\documentclass[superscriptaddress,nofootinbib,notitlepage,prx,aps,10pt,longbibliography,onecolumn]{revtex4-2}

\usepackage{fontspec} 
\usepackage{graphicx} 
\usepackage{hyperref} 
\hypersetup{
	pdfstartview={FitH},
	pdfnewwindow=true,
	colorlinks=true,
	linkcolor=blue,
	citecolor=blue,
	filecolor=blue,
	urlcolor=blue}
\usepackage{xcolor} 
\usepackage{amsmath} 
\usepackage{amssymb} 
\usepackage{amsfonts}
\usepackage{unicode-math} 
\usepackage{microtype}
\usepackage{float}
\usepackage[capitalize]{cleveref}

\usepackage{physics} 

\newcommand{\MQ}{\affiliation{%
School of Mathematical and Physical Sciences,
Macquarie University, Sydney, NSW 2109, Australia} }
\newcommand{\GU}{\affiliation{%
Centre for Quantum Dynamics,
Griffith University, Brisbane, Queensland 4111, Australia} }

\newcommand{\Sout}[1]{S_{#1,\beta}}
\newcommand{\nn}{\nonumber \\}
\newcommand{\coh}{\mathfrak{C}}

\newcommand{\Lorg}{{A}}
\newcommand{\Lorf}{{L}}

\begin{document}

\title{Cascading amplifiers can create exponentially large coherence}
\author{Lucas A. Ostrowski}\thanks{Both first authors contributed equally to the work.}\GU
\author{Giacomo Pantaleoni}\thanks{Both first authors contributed equally to the work.}\MQ
\author{Howard M. Wiseman}\GU
\author{Dominic W. Berry}\MQ

\begin{abstract} 
A standard laser beam has photon degeneracy, or coherence, $\mathfrak{C}$, of at most $8\mu^2$, where $\mu$ is the number of photons in the laser itself. Even quantum-engineered lasers, if required to produce a beam with the standard statistical properties, have limited coherence, scaling as $\mu^4$.  Moreover, such lasers (still unrealised) require very unconventional gain and output-coupling mechanisms. Here, we propose a different path to increasing $\mathfrak{C}$: cascaded linear amplifiers, with conventional couplings. By dropping the requirement on the beam’s properties, this approach can, in theory, achieve $\mathfrak{C}$ scaling \emph{exponentially} in $\mu$. Here $\mu$ is the total source excitation number, across all the amplifiers. Two amplifiers suffice to surpass the standard $\mu^2$ scaling. \end{abstract}
\maketitle

\tableofcontents

\section{Introduction}

Continuous-wave laser light is exceptional in its ability to exhibit a high degree of optical coherence~\cite{Glauber1963}, which can be highlighted through several characteristic features. Among these, a particularly distinctive feature of laser beams is the mean number of photons within the maximally populated mode of the light field, within a relevant frequency band\footnote{Consideration of a frequency band avoids issues associated with finite-temperature sources having a diverging mean photon number per mode at low frequencies. Although these modes have a high photon number, they are not at a useful frequency.}. In the early years of laser research, a quantity that was effectively the same as this was defined by Mandel~\cite{Mandel_1961}. He considered the number of photons in a ``coherence volume''~\cite{MW1995,Milonni_2010}, and referred to this as the \textit{photon degeneracy}. Even by 1961, lasers had been realised that produced photon degeneracies exceeding those from blackbody light sources by eleven orders of magnitude~\cite{Mandel_1961}, and this gap has only widened with more modern devices~\cite{Wiseman2016}. More recently, a more rigorous and generally applicable definition of this quantity was provided in Ref.~\cite{Baker2020} in terms of mode occupation, as given above. This quantity was denoted by $\mathfrak{C}$, and the term \textit{coherence} was coined for it.
For the case of a standard ideal laser beam, it does indeed equal (ignoring factors of order unity) the number of photons in a coherence length~\cite{Baker2020}. Thus, $\coh$ pinpoints a fundamental distinction between laser radiation and the emission from other common light sources.

The central result of Ref.~\cite{Baker2020} was to demonstrate the existence of an achievable upper bound for $\mathfrak{C}$ scaling as $\mu^4$, where $\mu$ is the resource that is used to produce $\mathfrak{C}$. This resource can often be regarded as the mean excitation number stored within the laser, but it is rigorously defined as the expectation value of a non-negative property of the device producing the beam (not of the beam), represented by operator $\hat n$, which acts as a generator of phase shifts on the beam. That is, applying the unitary operator $e^{i\theta\hat{n}}$ to the state of the laser device at time $t$ induces a relative phase shift of $\theta$ between the beam emitted prior to $t$ and the beam emitted subsequent to $t$.
The derivation of this upper bound assumed nothing of how the device operates, other than the following four conditions:
\begin{enumerate}
    \item \textbf{One Dimensional Beam}\label{cond:1} --- The device produces a single output beam, propagating with a fixed speed and direction, and occupying a single transverse mode with a single polarisation.
    \item \textbf{Endogenous Phase}\label{cond:2} --- Phase information proceeds only from the laser device, with no external phase injection. Rigorously, this condition is equivalent to the existence of the operator $\hat n$ defined above.
    \item \textbf{Stationarity}\label{cond:3} --- The long-time operation of the device is invariant under time-translation. This ensures that $\mu:=\langle\hat{n}\rangle$, the relevant resource for the production of $\mathfrak{C}$, has a unique value.
    \item \textbf{Glauber Ideality}\label{cond:4.0} --- The first- and second-order Glauber coherence functions of the beam well-approximate those of an ideal laser beam.
\end{enumerate}
Of these, the first three capture the essence of the concept of a source of a continuous-wave coherent beam, while the fourth condition stands out as rather restrictive. It imposes constraints on the beam that give rise to other prominent features of laser light, including a Lorentzian power spectrum and Poissonian photon statistics. 

As noted above, Ref.~\cite{Baker2020} not only proved that $\mathfrak{C}=O(\mu^4)$ (that is, an upper bound scaling), but also that there are quantum mechanical laser models, satisfying conditions \ref{cond:1}--\ref{cond:4.0}, that achieve this scaling: $\mathfrak{C} \propto \mu^4$. This result should be compared to standard ideal laser models, e.g.,~\cite{Louisell1973,SSL_1974}, comprising a single-mode slightly leaky cavity with a saturated gain process, which achieve $\mathfrak{C}=8\mu^2$ at best. This quadratic difference in scaling, for the same task, compelled Ref.~\cite{Baker2020} to identify $\mathfrak{C} \propto \mu^2$ as the standard quantum limit (SQL) scaling, and $\mathfrak{C} \propto \mu^4$ as the Heisenberg limit scaling for laser coherence.  A key feature of the models currently identified to surpass the SQL scaling is that they feature highly nonlinear input-output interactions with their environment~\cite{Liu2021,Baker2020,Ostrowski2022b}, effectively a form of quantum reservoir engineering~\cite{Poyatos_1996}.

Returning to the restrictiveness of the Glauber Ideality Condition, \ref{cond:4.0}, an obvious question is whether dropping this condition on the beam statistics would allow one to achieve 
the aforementioned $\mathfrak{C}\propto\mu^4$ scaling more easily, or even to surpass it. In this article, we answer this question positively: the scaling can, in fact, be arbitrarily surpassed, and with no requirement for nonlinear input-output interactions between the device and its environment. We demonstrate this by constructing an explicit physical model with the desired properties.
This model consists of a cascade of $N$ phase-insensitive single-mode linear amplifiers, where the output of each amplifier unidirectionally feeds into the one directly following it, while the first amplifier is fed by the vacuum field. 
We show that, with an optimized choice of the gain and bandwidth of each amplifier, one can obtain $\mathfrak{C}\propto(2\mu/N)^{2N}$ for fixed $N$. 
Moreover, if $N$ is allowed to depend on $\mu$, we show that the coherence can have, in principle, an {\em exponential} dependence on the photon number, $\mathfrak{C}\propto\exp(4\mu)$.

There are two important points about this coherence improvement, which we should make now, though we will also return to them throughout this work. The first is the proper accounting of the resource $\mu$, which, in our view, is often overlooked in studies on enhancing optical coherence in a light beam (see, e.g.,~\cite{Ritsch_1992,Kolobov_1993,van_Exter_1995,Loughlin_2023}). 
The Endogenous Phase Condition (Condition~\ref{cond:2}) requires a unique operator $\hat n$ with spectrum $\mathbb{N}_0$ that generates a phase shift on the future output beam. In our model, this operator is the sum of the photon number operators for each amplifier mode.
This makes sense because each of the $N$ amplifiers in our device stores coherent energy contributing to the increase of $\mathfrak{C}$, so all intra-amplifier excitations must be included in the total tally of $\mu$. 
It turns out that, in the optimal regime with large $\mu$, each successive amplifier in the cascade has a much larger bandwidth than the preceding one, but the number of photons in each one is the same, equal to $\mu/N$. In this regime, 
the beam photon flux is essentially the same as that of the final amplifier considered in isolation (reminiscent of an injection-locked laser~\cite{Nabors_1989}), but all amplifiers are needed to generate the coherence $\coh\propto (2\mu/N)^{2N}$.  

The second point we wish to emphasise is that our system substantially deviates from Glauber Ideality (Condition \ref{cond:4.0}). This is necessary to produce the improvement in the scaling beyond the Heisenberg limit $\mathfrak{C}\propto\mu^4$. Rather than a single coherence time for phase correlations as in an ideal laser beam~\cite{Louisell1973,SSL_1974}, our system has a hierarchy of coherence times, corresponding to the reciprocals of the bandwidths of the $N$ amplifiers. Specifically, Glauber's first-order coherence function~\cite{Glauber1963} is not the decaying exponential of an ideal laser, but a sum of differently weighted decaying exponentials. Moreover, the intensity fluctuations are also very different from those of an ideal laser beam, being much larger and also exhibiting multiple timescales. The overall larger intensity fluctuations are due to the unsaturated gain in each linear amplifier. We focus on this regime of linear gain because Gaussian methods can be applied, making the analysis far simpler than one involving above-threshold amplifiers. Intensity fluctuations influence both Glauber's first- and second-order coherence functions, and result in further deviation from Glauber Ideality. However, detailed analysis suggests that large intensity fluctuations do not play an essential role in our system's ability to achieve coherence beyond the SQL and Heisenberg limit.

The remainder of this article is structured as follows.  In Sec.~\ref{sec:model}, we introduce our system of $N$ cascaded linear amplifiers, along with its physical properties that are key to this work. In Sec.~\ref{sec:asymp}, we analyse the coherence of this system in an idealised setting and show that the $\mu^4$ Heisenberg scaling for coherence can be surpassed, even as far as $\exp(4\mu)$ in principle. In Sec.~\ref{sec:coherences}, we verify the consistency of this result by deriving a lower bound on the resource $\mu$, given $\mathfrak{C}$, for our system using methods from Ref.~\cite{Baker2020}. There, it is emphasised how this bound changes as a result of deviating from Glauber Ideality (Condition \ref{cond:4.0}).  In Sec.~\ref{sec:realistic}, we provide a more detailed analysis of our system in the $N=2$ case, and demonstrate that coherence beyond the $8\mu^2$ SQL can be achieved with realistic choices of values for the system parameters. In Sec.~\ref{sec:conc}, we conclude with a summary and a discussion. This includes conjecturing a more general (without Condition \ref{cond:4.0}) scaling law for the achievable upper bound on $\mathfrak{C}$, in terms of the dimensionality, rather than the mean excitation number $\mu$, of the device emitting the beam.

\section{Cascaded amplifiers}

\label{sec:model}

\subsection{Model}

\begin{figure}[H]
\centering
\includegraphics[width=0.8\columnwidth]{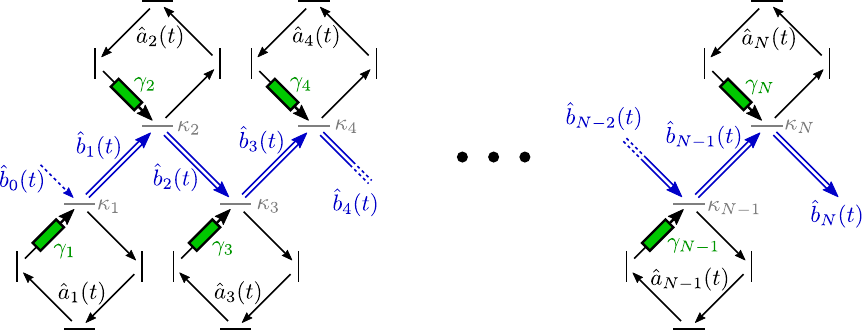}
\caption{\label{fig:1} A system of $N$ cascaded linear amplifiers. The $j$-th amplifier supports an internal mode with corresponding annihilation operator $\hat{a}_j(t)$, with a linear gain component of rate $\gamma_j$ and a loss of rate $\kappa_j$. The loss is responsible for the coupling between this internal field and the external input and output fields depicted with blue arrows. Respectively, these input and output fields have annihilation operators $\hat{b}_{j-1}(t)$ and $\hat{b}_{j}(t)$.
}
\end{figure}

In this section, we present a mathematical model for a system of $N$ linear amplifiers, which will be the subject of analysis in this work. 
All amplifiers have the same frequency and are cascaded in a unidirectional manner as depicted in Fig.~\ref{fig:1}. The input of the first is the vacuum, while the input of each successive amplifier is the output of the previous one. The $j$-th amplifier is a single-mode cavity, with output coupling rate $\kappa_j$, and containing an unsaturated gain medium characterised by a linear gain rate $\gamma_j$. It is assumed to be in the subthreshold regime\footnote{Note that this is distinct from the case of a laser with saturated gain, which is nonlinear gain. A single amplifier with fully saturated gain would give rise to a beam with
ideal Glauber coherence, which were the beam properties assumed in Ref.~\cite{Baker2020}.
See Chapter 9.3.4 of Ref.~\cite{Gardiner2004} for further explanation.} $\gamma_j < \kappa_j$. From the theory of cascaded quantum systems~\cite{Gardiner_1993,Carmichael_1993}, the corresponding annihilation operator $\hat{a}_j(t)$ for this mode evolves according to the quantum Langevin equation
\begin{align}
    \frac{d\hat{a}_j}{dt} = -\frac{1}{2}\left(\kappa_j - \gamma_j\right)\hat{a}_j(t) + \sqrt{\gamma_j}\hat{h}_{j}(t)-\sqrt{\kappa_j}\hat{b}_{j-1}(t),
\end{align}
where $\hat{h}_j(t)$ denotes the input field for the gain, which is modelled as an inverted bosonic heat bath at zero temperature so that $[\hat{h}^\dagger_j(t),\hat{h}_j(t')] = \delta(t-t')$ and $\langle\hat{h}^\dagger_j(t')\hat{h}_j(t)\rangle=\delta(t-t')$~\cite{Gardiner2004}. The output field from the $j$-th amplifier [which is the input field to the $(j+1)$-th one] is described by the annihilation operator $\hat{b}_j(t)$. This continuum-field operator has units of time$^{\rm -1/2}$ and satisfies the commutation relation $[\hat{b}_j(t),\hat{b}_j^\dagger(t')] = \delta(t-t')$. These output fields are related to one another, and the intracavity fields, via the usual input-output relation~\cite{Gardiner_1985} 
\begin{align}\label{eq:inout_general}
    \hat{b}_j(t) = \sqrt{\kappa_j}\hat{a}_j(t)+\hat{b}_{j-1}(t),
\end{align}
with $\langle\hat{b}^\dagger_0(t')\hat{b}_0(t)\rangle=0$. We are interested in the steady state of the above system, where all statistics are stationary. 

Given that the dynamics of our system are linear and the noise is quantum Gaussian~\cite{Gardiner_1985}, its state can be described in terms of an equivalent classical stochastic process for coherent states~\cite{Walls_Milburn_2008}. That is, we may express $\hat{b}_j(t)$ as a sum involving a stochastic coherent amplitude (a c-number) and the vacuum field $\hat{b}_0(t)$,
\begin{align}
    \hat{b}_j(t) = \beta_j(t) + \hat{b}_0(t).
\end{align}
Moreover, the internal cavity field operator $\hat{a}_j(t)$ may be replaced with a dimensionless complex amplitude $\alpha_j(t)$, which evolves according to the stochastic differential equation~\cite{Gardiner2004}
\begin{align}
\label{eq:langevint}
  d \alpha_j = - \frac{1}{2} \pqty{\kappa_{j} - \gamma_{j}} \alpha_j d t
  + \sqrt{\gamma_j} d V_j(t)
  - \sqrt{\kappa_j} \beta_{j - 1} dt
  \,.
\end{align}
The noise term is a complex Wiener increment associated with the \(j\)-th amplifier, where \(dV_j(t)=\eta_j(t)dt\) and $\eta_j(t)$ is complex Gaussian white noise. This increment has the moments $\langle dV_j(t)\rangle = \langle dV^*_j(t)\rangle = \langle dV_j(t)dV_j(t')\rangle = \langle dV^*_j(t)dV^*_j(t')\rangle = 0$ and $\langle dV^*_j(t)dV_j(t)\rangle = dt$. 
Below, we will show that all relevant properties of the beam can be deduced from the classical equations defined in Eq.~\eqref{eq:langevint} and the relations between the internal and external field amplitudes, which follow from Eq.~\eqref{eq:inout_general},
\begin{equation}\label{eq:inpout}
    \beta_j(t) = \sqrt{\kappa_j} \alpha_j(t) + \beta_{j-1}(t)\, .
\end{equation}

With this model, we now define some parameters which will be useful for interpreting our derived results. First, we define
\begin{align}\label{eq:def_ellj}
    \ell_j := \kappa_j-\gamma_j > 0,
\end{align}
which quantifies the bandwidth of the $j^{\rm th}$ amplifier. This can be thought of as the rate at which it relaxes to a steady-state mean amplitude if it were driven by a constant field [i.e., $\beta_{j-1}(t)=\beta_{j-1}$].  We will work in a regime where the amplitude $\beta_{j-1}$ varies adiabatically relative to $\alpha_j$ (this is where $\ell_{j-1}\ll\ell_j$, roughly speaking). This allows the approximation where $\beta_{j-1}$ can effectively be treated as a constant in Eq.~\eqref{eq:langevint}, which [ignoring noise $dV_j(t)$ for the purposes of this heuristic argument] yields a quasi-steady state for the mean amplitude of $\alpha_j\approx-2\sqrt{\kappa_j}\beta_{j-1}/(\kappa_j-\gamma_j)$. Equation~\eqref{eq:inpout} then gives $\beta_j\approx-(\kappa_j+\gamma_j)\beta_{j-1}/(\kappa_j-\gamma_j)$, implying an amplification of the mean field $\beta_{j-1}$ as it passes through amplifier $j$. In anticipation of this regime, we define the dimensionless quantities
\begin{align}\label{eq:def_Gj}
    G_j := \left(\frac{\kappa_j+\gamma_j}{\kappa_j-\gamma_j}\right)^2 > 1,
\end{align}
which can be interpreted as the photon number gain~\cite{Caves_1982,Clerk_2010}, or simply the gain, of the input field $\beta_{j-1}$ by amplifier $j$. Note that \cref{eq:def_Gj} is a definition, not an approximation. The approximations discussed in this paragraph are merely to motivate the definition.

\subsection{Coherence and mean photon number}

We wish to determine the relationship between coherence of the output, from the final amplifier in the cascade, and the total internal photon number of the system. With our formulation, we have implicitly assumed a frame that is rotating with the frequency of each cavity; from this property, and the time-translational invariance of the system, the coherence of the $N$-th output may be expressed as~\cite{Baker2020}
\begin{align}\label{eq:coh_g1}
    \mathfrak{C}_N = \mathcal{N}_N\int_{-\infty}^{\infty}d\tau\,g^{(1)}_N(t+\tau,t).
\end{align}
Here $\mathcal{N}_j:=\langle\hat{b}_j^\dagger(t)\hat{b}_j(t)\rangle$ is the photon flux of the $j$-th amplifier and we are also using the (normalised) first-order Glauber coherence function for the $N$-th output $g^{(1)}_N(t+\tau,t)=\langle\hat{b}_N^\dagger(t+\tau)\hat{b}_N(t)\rangle/\mathcal{N}_N$. More generally, we can define a family of $n$-th order Glauber coherence functions~\cite{Glauber1963} involving $2n$ normally-ordered field operators
\begin{align}\label{eq:gn_general}    g^{(n)}_j(s_1,\dots,s_{2n}):=\frac{\langle\hat{b}_j^\dagger(s_1)\cdots\hat{b}_j^\dagger(s_n)\hat{b}_j(s_{n+1})\cdots\hat{b}_j(s_{2n})\rangle}{\prod_{i=1}^{2n}\left[\langle\hat{b}_j^\dagger(s_i)\hat{b}_j(s_{i})\rangle\right]^{1/2}} = \frac{\langle\beta_j^*(s_1)\cdots\beta_j^*(s_n)\beta_j(s_{n+1})\cdots\beta_j(s_{2n})\rangle}{\prod_{i=1}^{2n}\left[\langle|\beta_j(s_i)|^2\rangle\right]^{1/2}},
\end{align}
which are useful for determining additional properties of the output beam from the $j$-th amplifier. In writing the equality on the right hand side of \cref{eq:gn_general}, we have used the fact that the vacuum fields will not contribute to normally-ordered operator products, so we can safely make the substitutions $\hat{b}_j(t)\to\beta_j(t)$ and $\hat{b}^\dagger_j(t)\to\beta_j^*(t)$. Similarly, the mean internal photon number of the system can be expressed in terms of the internal complex amplitudes,
\begin{align}\label{eq:mu_general}
    \mu= \sum_{j=1}^N\mu_j, \quad \quad {\rm with} \quad \quad\mu_j:= \langle\hat{a}^\dagger_j(t)\hat{a}_j(t)\rangle = \langle|\alpha_j(t)|^2\rangle .
\end{align}
To be precise, photons in free-space propagation between amplifiers should also be counted alongside the intra-amplifier photons in the total tally of $\mu$ to satisfy Condition~\ref{cond:2}. However, these free-space photons can be made an arbitrarily small fraction of the total by ensuring that the loss rates in each amplifier are sufficiently small, where $\kappa_jt_j\ll1$ and $t_j$ is the propagation time of light from amplifier $j$ to $j+1$. Our calculations assume this regime.

We can obtain expressions for $\mathfrak{C}_N$ and $\mu$ in terms of the system parameters $\gamma_j$ and $\kappa_j$ by working in the frequency domain. For this purpose, we let $\widetilde\Sout{j}(\omega)$ and $\widetilde S_{j,\alpha}(\omega)$ denote the output and intracavity power spectra for the $j$-th amplifier, respectively. These are related to the correlation functions of the internal and external field amplitudes via Fourier transforms
\begin{align}\label{eq:spdef}
	\delta(\omega - \omega') \widetilde\Sout{j}(\omega) &= \langle \tilde \beta^{*}_{j} (\omega) \tilde \beta_{j} (\omega') \rangle
	\, , \\ \label{eq:spdef_2}
\delta(\omega - \omega') \widetilde S_{j,\alpha}(\omega) &= \langle \tilde \alpha^*_j(\omega) \widetilde \alpha_j(\omega') \rangle \, ,
\end{align}
where the convention for the Fourier transform of an arbitrary function $f(t)$ that we use is $\tilde{f}(\omega):= (1/2\pi)\int_{-\infty}^\infty d\tau\,f(\tau)e^{i\omega \tau}$ (following Ref.~\cite{MW1995}). The coherence and total mean cavity photon number may then be determined in terms of these power spectra as
\begin{align}
    \label{eq:coh_spec}
    \mathfrak{C}_{N} & = 2\pi\widetilde\Sout{N}(0), \\
    \label{eq:mu_spec}
    \mu  = & \int d \omega \sum_{j = 1}^N \widetilde S_{j,\alpha}(\omega).
\end{align}

To evaluate Eqs.~\eqref{eq:coh_spec} and \eqref{eq:mu_spec} for our model, we take the Fourier transform of \cref{eq:langevint} and find
\begin{align}
\label{eq:langevinomegaalpha}
  \tilde \alpha_j (\omega)
  &=
    - \frac{2}{\pqty{\gamma_j - \kappa_j + 2 i \omega}} \pqty{
    \sqrt{\gamma_j} \tilde \eta_j (\omega) - \sqrt{\kappa_j} \tilde \beta_{j - 1} (\omega)}
    \, ,
\end{align}
where $\tilde \eta_{j}(\omega)=(1/2\pi)\int_{-\infty}^\infty dV_j(\tau)e^{i\omega\tau}$.
The random processes in each cavity are independent, so their correlations are $\langle \tilde \eta_j^*(\omega) \tilde \eta_k (\omega') \rangle = \delta_{jk}\delta(\omega - \omega')/2\pi$.
It is also useful to write the Langevin equations in terms of \(\tilde \beta_{j}\) rather than \(\tilde \alpha_{j}\) by making use of the input-output relations in \cref{eq:inpout} to give
\begin{align}
    \label{eq:derivation0}
    \tilde \beta_j(\omega)
    &= \Lorg_j(\omega) \tilde \eta_j (\omega) + \tilde \beta_{j - 1} (\omega) \pqty{1 - \sqrt{\frac{\kappa_j}{\gamma_j}} \Lorg_j (\omega)}
    \,,
\end{align}
where we have defined
\begin{align} \label{eq:gdefinition}
  \Lorg_j(\omega) := - \frac{2 \sqrt{\gamma_{j} \kappa_{j}}}{\gamma_{j} - \kappa_{j} + 2 i \omega}
  \, .
\end{align}
Note that the square modulus of $\Lorg_j$ is a Lorentzian function.

The power spectrum from the $j$-th cavity (and hence the coherence of its output field) can be obtained from the correlation function $\langle \tilde \beta^*_j (\omega) \tilde \beta_j (\omega')\rangle$, which can be evaluated by applying \cref{eq:derivation0} recursively. At the very first step, the correlation $\langle \tilde \beta^*_j (\omega) \tilde \beta_j (\omega')\rangle$ becomes a sum of four separate correlations involving stochastic averages of products of $\tilde \eta_j$, $\tilde \eta^*_j$, $\tilde \beta_{j-1}$, and $\tilde \beta^*_{j-1}$. The first term in the sum is proportional to $\langle \tilde \eta^*_j (\omega) \tilde \eta_j (\omega')\rangle = \delta(\omega-\omega')/2\pi$, while the cross terms are the stochastic averages of \(\tilde \eta^{*}_{j} (\omega) \tilde \beta_{j-1} (\omega')\) and its complex conjugate. Since the $j$-th stochastic process is uncorrelated with the $(j-1)$-th output field upstream, these cross terms are exactly zero. One may verify this causality property by noting that these averages are sums of vanishing Kronecker deltas. Thus, only two terms contribute, which gives us
\begin{align}
\label{eq:derivation1}
  &\langle \tilde \beta^*_j (\omega) \tilde \beta_j (\omega')\rangle =
  \abs{\Lorg_j(\omega)}^{2} \frac{\delta(\omega - \omega')}{2 \pi} +
  \pqty{1 - \sqrt{\frac{\kappa_j}{\gamma_j}} \Lorg^*_j(\omega)}
  \pqty{1 - \sqrt{\frac{\kappa_j}{\gamma_j}} \Lorg_j(\omega')}
  \langle \tilde \beta^*_{j - 1}(\omega) \tilde \beta_{j - 1}(\omega') \rangle
  \,.
\end{align}
Using \cref{eq:spdef} on both sides of Eq.~\eqref{eq:derivation1}, we obtain factors multiplying Dirac delta distributions on either side.
Factoring out these delta distributions gives a recursion relation that involves the power spectra more explicitly
\begin{align}\label{S_N,out}
  \widetilde \Sout{j}(\omega) = \frac{\Lorf_j(\omega)}{2\pi} + \pqty{1 + \Lorf_j(\omega)} \widetilde \Sout{j - 1}(\omega) \,, \quad \Sout{0} = 0
  \,,
\end{align}
where we have defined the Lorentzian-shaped functions
\begin{equation}\label{eq:fdef}
    \Lorf_j(\omega) := \abs{\Lorg_{j}(\omega)}^2 = \frac {\gamma_j\kappa_j}{\left(\frac{\gamma_j-\kappa_j}{2}\right)^2+\omega^2} = \frac{(\ell_j/2)^2\left(G_j-1\right)}{\left(\ell_j/2\right)^2+\omega^2}\, .
\end{equation}
The iterative formula~\eqref{S_N,out} yields the closed form solution
\begin{align}\label{S_N,out_2}
  2\pi \widetilde \Sout{j}(\omega) &= \prod_{k = 1}^j (1 + \Lorf_k(\omega)) - 1 \,.
\end{align}
Using this with \cref{eq:coh_spec} gives us
\begin{align}
\label{eq:coherencen}
\mathfrak{C}_{N} & =
\prod_{j = 1}^N (1 + \Lorf_j(0)) - 1 = \prod_{j = 1}^N G_j - 1 \,.
\end{align}

For the photon number inside the amplifiers, we need the intracavity power spectrum $\widetilde S_{j,\alpha}(\omega)$. This quantity can be obtained from Eqs.~\eqref{eq:spdef_2} and \eqref{eq:langevinomegaalpha},  
\begin{align}
\label{eq:essjay}
   \widetilde S_{j,\alpha}(\omega) = \frac{\Lorf_j(\omega)}{2 \pi \kappa_j} + \frac{\Lorf_j(\omega)\widetilde \Sout{j - 1}(\omega)}{\gamma_j}
  \, .
\end{align}
Using this with \cref{eq:mu_spec}, the total photon number for the system of $N$ cascaded amplifiers is therefore
\begin{align} \label{eq:photonnumber}
  \mu
  &= \int d \omega \sum_{j = 1}^N \Lorf_j(\omega)\left[\frac{1}{ 2 \pi \kappa_j} + \frac{\widetilde \Sout{j - 1}(\omega)}{\gamma_j} \right]
\,.
\end{align}

\section{Asymptotic results for the coherence} 
\label{sec:asymp}

We now determine the asymptotic scaling of the coherence with the total intracavity photon number for our system using a simplified choice of parameter values. We show that using two amplifiers enables the coherence to scale at the Heisenberg limit from Ref.~\cite{Baker2020}, and that increasing the number of amplifiers increases the scaling without limit, ultimately leading to an exponential dependence of the coherence in terms of the intracavity photon number.

The simplified choice of parameter values that we consider is
\begin{align}\label{eq:ncavityconfig}
  &\ell_{j} = r^{j-1}\ell_1, 
  \\ \label{eq:ncavityconfig2}
  &G_j  = G.
\end{align}
This sets a fixed ratio $r=\ell_j/\ell_{j-1}$ between the bandwidths of successive amplifiers, along with a fixed photon number gain $G$ for each amplifier. With this restriction, our system is thus characterised in terms of four parameters: one rate $\ell_1$, and three dimensionless quantities $r$, $G$, and $N$.

Using this parameter configuration, Eq.~\eqref{eq:coherencen} becomes
\begin{align}\label{eq:CN}
\mathfrak{C}_{N} &= G^{N} -1 \,.
\end{align}
It turns out that the asymptotic scaling of this quantity with $\mu$ is most favourable in the \textit{large $r$ regime}, by which we mean the limit $G\ll r$. We thus restrict ourselves to this regime in the remainder of this section, 
obtaining results which are independent of $r$.
Physically, having $r$ large corresponds to the scenario where the evolution timescale of each cavity is much faster than the previous one. In practice, $r$ could not be too large, but taking more moderate values of this parameter considerably complicates the analysis. We address this issue in Section \ref{sec:realistic}.

In Appendix~\ref{app:derive_g1}, we show that the large $r$ regime is optimal, and that, in this regime, the total photon number is the sum of the intracavity photon numbers for each amplifier as if they were uncoupled. Explicitly, the mean excitation number for each cavity asymptotically equals $\frac12\big(\sqrt{G}-1\big)$, giving 
\begin{align}\label{eq:mu_sim}
  \mu \underset{r}{\sim} N\frac{\sqrt{G}-1}{2}.
\end{align}
Here, the ``$\underset{r}{\sim}$'' symbol means equal up to quantities that are 
vanishingly small as $G/r\to0$. An expression for the coherence as a function of photon number can then be found by inverting \cref{eq:mu_sim} with respect to $G$ and substituting $G(\mu)$ into \cref{eq:CN}. This yields
\begin{align}\label{eq:coherenceasymn}
	\mathfrak{C}_{N} \underset{r}{\sim}
        \left( 1 + \frac{2 \mu}{N} \right)^{2N}-1
\, .
\end{align}

Now, there are two obvious ways to analyse the asymptotic scaling of \cref{eq:coherenceasymn} with $\mu$.
The first is to consider $N$ fixed. Then the large $\mu$ limit requires 
the intracavity photon number for each amplifier to be large. 
This in turn requires $\sqrt{G}$ to be large, which we refer to as the \textit{large gain regime}. 
For this to be compatible with the large $r$ regime, we need 
\begin{equation} \label{bothlarge}
    1\ll G \ll r. 
\end{equation}
If this condition is met, then \cref{eq:coherenceasymn} simplifies to 
\begin{align}\label{eq:asympCN}
\mathfrak{C}_{N} \underset{r,\mu}{\sim} 
\frac{\mu^{2N}}{(N/2)^{2N}} \, ,
\end{align}
where $\underset{r,\mu}{\sim}$ means the asymptotic behaviour of $\mathfrak{C}$ as $\mu\to\infty$, where $r$ is allowed to depend on the target $\mu$ to maintain $r\gg G$. This results from the fact that, in this regime, $\mathfrak{C}_N \underset{r,\mu}{\sim} G^N$ while $\mu \underset{r,\mu}{\sim} N\sqrt{G}/2$.

\Cref{eq:asympCN} reveals that, by increasing the number of cavities, it is possible to obtain a coherence scaling as an arbitrarily large power of $\mu$, albeit with a prefactor $1/(N/2)^{2N}$ that becomes smaller for larger $N$. 
For $N>2$, the coherence is scaling as a higher power of $\mu$ than 
the $\mu^4$ Heisenberg limit for laser coherence from Ref.~\cite{Baker2020}. The latter is the \textit{ultimate} quantum limit for $\mathfrak{C}$ when the Glauber coherence functions of the beam satisfy Glauber Ideality (Condition \ref{cond:4.0}) mentioned in the introduction.
This implies that the Glauber coherence functions of our model here must deviate considerably from those of an ideal laser beam. In Sec.~\ref{sec:coherences}, we address this point in detail.

The second way to analyse the asymptotic scaling of \cref{eq:coherenceasymn} with $\mu$ is to let $N$ depend on $\mu$, which itself can be done in different ways. In the limit that $N\gg\mu^2$ and $\mu$ becomes large,~\cref{eq:coherenceasymn} becomes
\begin{equation}\label{eq:exp_CN}
    \mathfrak{C}_{N\gg\mu^2} \underset{r,\mu}{\sim} e^{4\mu} \, .
\end{equation}
This formula should be understood as the asymptotic scaling of $\mathfrak{C}_N$ with $\mu$, as the number of amplifiers scales (at least) as the square of the mean photon number. Note that, in this case, $r$ can be considered a fixed large number because $G$ does not increase with $\mu$. Equation \eqref{eq:exp_CN} establishes the exponential dependence of the coherence on $\mu$, which is the result given in the title. It is worth highlighting, however, that the asymptotic expression in Eq.~\eqref{eq:exp_CN} would not be practical to achieve experimentally. This is because, with $r\gg 1$, the smallest linewidth in the cascade, $\ell_1 = \ell_N r^{1-N}$, is obviously unphysically small if $N\gg\mu^2\gg1$, since the largest linewidth must satisfy $\ell_N \ll \omega_0$, the central frequency of the radiation. (Any choice of parameters satisfying these inequalities gives $1/\ell_1$ far longer than the lifetime of the universe.)

Even though Eq.~\eqref{eq:exp_CN} cannot feasibly be attained, exponential dependence of the coherence on the mean photon number may still be observed for moderate $\mu$. To see this, suppose that $G$ is held fixed, so that the mean photon number in each amplifier $\mu_j=\mu/N$ is constant [from Eq.~\eqref{eq:mu_sim}, we know that $\mu_j=(\sqrt G-1)/2$]. Then, the total mean photon number can be varied (in discrete steps) by varying the number of amplifiers. Rewriting~\cref{eq:coherenceasymn} in terms of $\mu$ and the constant $\mu_1 = \mu/N$ gives exactly
\begin{align}\label{eq:coh_exp}
    \mathfrak{C}_{N = \mu/\mu_j} \underset{r}{\sim} \exp\left[\left\{\frac{\ln\left(1+2\mu_1\right)}{2\mu_1}\right\}4\mu\right]-1.
\end{align}
This formula should be understood as the behaviour of $\mathfrak{C}_N$ as the mean photon number scales proportional to the number of amplifiers, for fixed $G$ and fixed $r\gg G$. We also emphasise that $\mu$ in this formula should be interpreted as a discrete variable, being an integer multiple of $\mu_1 = (\sqrt G-1)/2$. It can be seen that the optimal choice for maximising the coherence in \cref{eq:coh_exp} would be to have $\mu_1\ll 1$, in which case the prefactor in the curly braces can be approximated by $1$.

\begin{figure}[H]
\centering
\includegraphics[width=0.5\columnwidth]{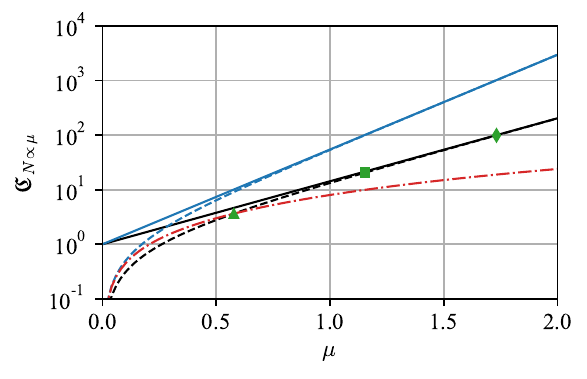}
\caption{\label{fig:exp_dep}Coherence, given by Eq.~\eqref{eq:coh_exp}, plotted as a function of $\mu$ for a $N=1$ (triangle), $N=2$ (square), and $N=3$ (diamond) amplifier chain, with $\mu/N = 1/\sqrt{3}$ photons in each amplifier. Black-dashed and black-solid curves, which plot \cref{eq:coh_exp} and the exponential part of \cref{eq:coh_exp} respectively, are guides for the eye, as they take $\mu$ to be a continuous variable with $\mu/N = 1/\sqrt{3}$. Blue-dashed and  blue-solid curves are the same as the respective black curves, but instead consider the optimal limit $\mu/N\ll1/\mu$, so that \cref{eq:coh_exp} is well-approximated by $\mathfrak{C}_{N\propto\mu} \underset{r}{\sim} \exp\left(4\mu\right)-1$. For completeness, the red-dash-dotted curve shows the coherence of a single linear amplifier as a function of $\mu$ [{\em i.e.}, $4\mu(\mu+1)$, from \cref{eq:coherenceasymn} with $N=1$].}
\end{figure}

A practical scenario that could verify the behaviour predicted in \cref{eq:coh_exp} would be to set up a cascade of three amplifiers operating in the large $r$ regime, and then individually measure the coherence of the output from the first, second, and third amplifiers, along with their respective photon numbers. Each measurement of the coherence could therefore be regarded as arising from an $N=1$, $N=2$, and $N=3$ amplifier chain, respectively. For instance, suppose each amplifier contains an average of $\mu_1=1/\sqrt{3}$ photons (this would require $G\approx4.643$). In this case, the prefactor in the curly braces of \cref{eq:coh_exp} is $\sqrt{3}\ln(1+2/\sqrt{3})/2\approx0.6648$, giving an exponential rate with $\mu$ of approximately $2.66$. The total mean photon number would be $\mu=1/\sqrt{3}$, $\mu=2/\sqrt{3}$, and $\mu=\sqrt{3}$ when considering the coherence of the output from the first ($N=1$), second ($N=2$), and third ($N=3$) amplifiers, respectively. One could then ideally obtain a plot similar to that shown in Fig.~\ref{fig:exp_dep}, revealing a coherence that approximately follows an exponential curve.
An important consideration here is the linewidth ratio of the third and first amplifiers $\ell_3/\ell_1=r^{2}$, which shouldn't be too large. If we assume that $G/r=0.1$ is sufficient to be within the large $r$ regime, then the parameter values above would give a linewidth ratio of $\ell_3/\ell_1 \approx 2155$, which is not unphysical.

\section{How the Glauber coherence functions influence the upper bounds for \texorpdfstring{$\mathfrak{C}$}{C}}
\label{sec:coherences}

In Ref.~\cite{Baker2020}, an upper bound for $\mathfrak{C}$ was derived for asymptotically large $\mu$, which we reproduce here as a lower bound on $\mu$ for asymptotically large $\mathfrak{C}$: 
\begin{align}\label{eq:baker_ub}
    \mu\gtrsim\left(\frac{3}{2}\right)^{1/4}\left|\frac{z_A}{3}\right|^{3/2}\mathfrak{C}^{1/4}.
\end{align}
Here $z_A\approx-2.338$ is the first zero of the Airy function and the $\gtrsim$ means greater than or equal up to vanishing contributions for large $\mu$. We express this as a lower bound on $\mu$ because the key aspect of the system we modify to improve coherence is the Glauber coherence functions of the beam. In the proof of this bound (see below), these coherence functions influence the behaviour of $\mathfrak{C}$ as opposed to $\mu$. 

Our system can achieve a relationship between $\mu$ and $\mathfrak{C}$ that breaks the bound in Eq.~\eqref{eq:baker_ub}. For instance, in the large gain regime where \cref{bothlarge} is fulfilled, Eq.~\eqref{eq:asympCN} yields
\begin{align}\label{eq:asym_mu}
    \mu \underset{r,\mu}{\sim} \frac{N}{2}\left(\mathfrak{C}_{N}\right)^{1/2N}.
\end{align}
For $N\geq 2$, this value of $\mu$ is clearly smaller than the lower bound given by Eq.~\eqref{eq:baker_ub}. There is no contradiction here because our system violates the Glauber Ideality Condition (Condition \ref{cond:4.0}) that was used to derive that bound; consequently, Eq.~\eqref{eq:baker_ub} does not represent a Heisenberg limit for our system. The purpose of this section is to explore how the beam of our system differs from that produced by an ideal laser, and why that difference enables the mean source excitation number, $\mu$, to be much smaller for a given beam coherence, $\coh$. To simplify our analysis, we restrict our attention in this section to a beam exhibiting the same statistics as our cascaded amplifier model in the large-gain regime.

\subsection{Comparison with an ideal laser beam}

An ideal laser beam state can be regarded as a coherent state that is undergoing a process of pure phase diffusion. If we let $\ell_{\rm ideal}$ denote the diffusion rate, and $\mathcal{N}_{\rm ideal}$ the photon flux, then this can be expressed as an eigenstate of $\hat{b}(t)$, with the stochastic eigenvalue $\sqrt{\mathcal{N}_{\rm ideal}}e^{i\sqrt{\ell_{\rm ideal}}W(t)}$, where $W(t)$ is a Wiener process. The first- and second-order Glauber coherence functions for this state are~\cite{Baker2020}
\begin{align}\label{eq:id_g1}
    g^{(1)}_{\rm ideal}(s,t) = \frac{\langle \hat{b}^\dagger(s)\hat{b}(t) \rangle_{\rm ideal}}{\mathcal{N}_{\rm ideal}} = e^{-\ell_{\rm ideal}|s-t|/2},
\end{align}
\begin{align}\label{g2_ideal}
    g^{(2)}_{\rm ideal}(s,s',t',t) = \frac{\langle \hat{b}^\dagger(s)\hat{b}^\dagger(s')\hat{b}(t')\hat{b}(t) \rangle_{\rm ideal}}{\mathcal{N}_{\rm ideal}^2} = \frac{e^{-\ell_{\rm ideal}\left(|s-t|+|s'-t'|+|s-t'|+|t-s'|\right)/2}}{e^{-\ell_{\rm ideal}\left(|s-s'|+|t-t'|\right)/2}}.
\end{align}
In order for Eq.~\eqref{eq:baker_ub} to represent a Heisenberg limit, it is necessary that the beam well-approximate these functions (in a specific sense; see Ref.~\cite{Baker2020}).

In Appendix~\ref{app:derive_g1}, we show that the first-order Glauber coherence function for our system can be approximated by
\begin{align}\label{g1_approx}
    g_N^{(1)}(s,t) \approx \sum_{j=1}^{N}\left(\frac{G}{r}\right)^{N-j}e^{-r^{j-1}\ell_1|s-t|/2},
\end{align}
while the photon flux is effectively the same as what it would be for the $N$-th amplifier with vacuum input
\begin{align}\label{flux_apx}
    \mathcal{N}_N \approx r^{N-1}\frac{(G-1)\ell_1}{4}.
\end{align}
An obvious distinction between $g_N^{(1)}$ and $g^{(1)}_{\rm ideal}$ is the multiple timescales of $g^{(1)}_N$, where the $j$-th term in the sum constituting  Eq.~\eqref{g1_approx} exhibits a decay rate of $r^{j-1}\ell_1$, intrinsic to the $j$-th amplifier. For times $(r^k\ell_1)^{-1}\ll|s-t|\ll(r^{k-1}\ell_1)^{-1}$, the $j=k$ term in the sum dominates, so the phase correlations of the beam predominantly arise from the $k$-th amplifier. Essentially, this means that there is not a single coherence time that can be associated with the beam as in the ideal laser beam; instead, phase correlations emerge and subsequently decay once for every amplifier within the cascade.

The second-order Glauber coherence function can be readily obtained from $g^{(1)}_N(s,t)$ by exploiting the Gaussian properties of the system state~\cite[Chapter 4.4.5]{Gardiner2004}
\begin{align}\label{g2_exact}
    g_N^{(2)}(s,s',t',t) = g_N^{(1)}(s,t)g_N^{(1)}(s',t') + g_N^{(1)}(s,t')g_N^{(1)}(s',t) \, .
\end{align}
It is clear from this expression that $g^{(2)}_N$ is also quite distinct from $g^{(2)}_{\rm ideal}$. There are multiple timescales associated with $g^{(2)}_N$, along with substantial intensity fluctuations. These intensity fluctuations can be highlighted if we evaluate the photon statistics of the beam at zero time delay, for which $g^{(2)}_N(t,t,t,t) = 2$. This correlation value implies photon bunching, and is more typical of thermal light rather than an ideal laser beam, which instead has uncorrelated photon emission, with $g^{(2)}_{\rm ideal}(t,t,t,t) = 1$. Such behaviour for $g^{(2)}_N$ is expected for linear amplifiers. It may be possible to reproduce a beam with a  $g^{(1)}$ (and hence $\coh$) similar to \cref{g1_approx}, but with intensity statistics more typical of a laser beam, by considering cascaded saturated amplifiers. However, as noted earlier, that would be far more difficult to model because it would be nonlinear and it would not be possible to simply linearise around a mean field because of the multiple timescales.

\subsection{Modified Glauber coherence yields a different upper bound on \texorpdfstring{$\mathfrak{C}$}{C}}

The Glauber coherence functions given in Eqs.~\eqref{g1_approx} and \eqref{g2_exact}, along with the photon flux in Eq.~\eqref{flux_apx}, significantly alter the lower bound for the internal mean photon number $\mu$ required to achieve a given coherence $\mathfrak{C}$. To understand these fundamental limits, we now abstract away from the specific cascaded amplifier model. We instead consider an \textit{arbitrary} quantum device that produces an output beam possessing these exact multi-timescale statistics, and ask: what is the absolute minimum $\mu$ that must be stored within such a device?

The derivation of this bound relies on the principle that the coherence $\mathfrak{C}$ depends entirely on the Glauber coherence properties of the emitted beam, regardless of the physical hardware used to generate it. However, producing a beam with these specific multi-timescale correlations inherently dictates a minimum required $\mu$ inside the device. We establish this relation by adapting the method from Ref.~\cite{Baker2020} and replacing their ideal laser statistics with the non-ideal beam statistics derived above. This yields a new lower bound on $\mu$ as a function of $\mathfrak{C}$ (or conversely, an upper bound on $\mathfrak{C}$ as a function of $\mu$). The argument is outlined below, with full details provided in Appendix~\ref{app:bounds}.

The bound between $\mu$ and $\mathfrak{C}$ can be obtained by considering the following scenario: Initially, an observer performs a heterodyne measurement on the beam for some time $\tau$, from which a phase estimate $\varphi_F$ may be extracted. Because of the Endogenous Phase Condition (Condition \ref{cond:2}), this imprints $\varphi_F$ onto the phase of the state stored in the device. Then, a second observer is tasked with estimating that imprinted device phase. This is done either by performing a canonical phase measurement directly on the device to obtain the phase estimate $\varphi_D$, or by letting the phase information leak out of the device and again performing heterodyne detection, on the newly generated beam, for a time $\tau$, to obtain the phase estimate $\varphi_R$. Condition \ref{cond:2}, together with the optimality of $\varphi_D$, implies that $\varphi_D$ necessarily performs at least as well as $\varphi_R$ as an estimate of $\varphi_F$. This is because the process of beam formation followed by a suboptimal measurement can only add noise to the internal phase information. Thus, we can assert an inequality between the circular variance of the estimates
\begin{align}\label{eq:MSE1}
    1-\left|\left\langle e^{i\left(\hat{\varphi}_F-\hat{\varphi}_D\right)}\right\rangle\right|^2 \leq 1-\left|\left\langle e^{i\left(\hat{\varphi}_F-\hat{\varphi}_R\right)}\right\rangle\right|^2.
\end{align}

The circular variance of the canonical measurement can be bounded below in terms of $\mu$ using a result from Ref.~\cite{Bandilla_1991} (which was derived for single-mode fields, but generalised to arbitrary systems in Ref.~\cite{Berry_2012}). This result states that, given an arbitrary state $\rho$ with mean excitation number $\mu$, the \textit{minimal} phase uncertainty for such a state is bounded below by the quantity $ 4\left|z_A/3\right|^3\mu^{-2}$, for asymptotically large $\mu$. Hence, the MSE in the canonical measurement must be at least as large as this minimal phase uncertainty; using this fact in Eq.~\eqref{eq:MSE1} yields
\begin{align}\label{eq:MSE2}
    4\left|\frac{z_A}{3}\right|^3\mu^{-2} \lesssim  1-\left|\left\langle e^{i\left(\hat{\varphi}_F-\hat{\varphi}_R\right)}\right\rangle\right|^2.
\end{align}

Up to this point, the derivation follows exactly that of Ref.~\cite{Baker2020}. A different bound on $\mu$ is obtained for our system due to its Glauber coherence functions modifying the behaviour of two main quantities:
\begin{enumerate}
    \item the coherence $\mathfrak{C}$, and
    \item the error in the phase estimate $\hat{\varphi}_R$ given in the right-hand side of \cref{eq:MSE2}.
\end{enumerate}
In particular, the coherence depends on $g_N^{(1)}$ and the error in the phase measurement depends on $g_N^{(2)}$, so changing the Glauber coherence from that for an ideal laser may change both. Note that these quantities depend entirely on the properties of the output beam, and it is Eq.~\eqref{eq:MSE2} that then bounds the properties of the generic device that could produce this beam.

Considering first the coherence $\mathfrak{C}$, this is obtained by integrating $g_N^{(1)}$ in accordance with \cref{eq:coh_g1}. In the large $r$ and large gain regime, $g_N^{(1)}$ is given by  \cref{g1_approx}, which has magnitudes scaling as $(r/G)^j$, but the exponential has $r^{j-1}$, cancelling the $r^j$ scaling when integrating over all times. That is, we obtain
\begin{align}
    \mathfrak{C}_N &= \mathcal{N}_N\int d\tau \, g_N^{(1)}(t+\tau,t)  \nn
    &= \mathcal{N}_N\int d\tau \sum_{j=1}^{N}\left(\frac{G}{r}\right)^{N-j}e^{-r^{j-1}\ell_1|\tau|/2} \label{sumofexpsint} \\
    &= \frac{4\mathcal{N}_N}{\ell_1}\frac{1}{r^{N-1}}\sum_{j=1}^{N}G^{N-j}
    \, . \nonumber
\end{align}
Since we are interested in the large $G$ regime, $G\gg 1$, the terms in the sum decrease geometrically with $j$, and the dominant contribution is that from $j=1$. Indeed, keeping only this term in the sum, and using $\mathcal{N}_N\sim r^{N-1}(G-1)\ell_1/4$ [\cref{flux_apx}],  yields
\begin{equation}
    \mathfrak{C}_N \approx  4 \mathcal{N}_N  \frac 1{\ell_1}\left(\frac{G}{r}\right)^{N-1} \sim G^{N}\, ,
\end{equation}
which agrees with the exact result \eqref{eq:CN} in the same large gain limit.  The point of this calculation is that, although the exponential corresponding to the slowest decay timescale ($j=1$) is the only one that contributes significantly to the integral in \cref{sumofexpsint}, the large final coherence results from this term being multiplied in this equation by the overall flux parameter, $\mathcal{N}_N$. This overall flux is, in this limit, much larger than the effective flux contribution from the slowest timescale because it is dominated by the fastest timescale dynamics.

Second, Ref.~\cite{Baker2020} derives a formula for the mean-square error (MSE) $\langle[\hat{\varphi}_F-\hat{\varphi}_R]^2\rangle$
\begin{align}\label{eq:mse}
    \frac 1{\mathcal{N}\tau} &+ \frac 1{2\tau^4} \left[ \int_0^\tau ds \int_{-\tau}^0 ds' \int_0^\tau dt' \int_{-\tau}^0 dt \, g^{(2)}(s,s',t',t) - \int_0^\tau ds \int_0^{\tau} ds' \int_{-\tau}^0 dt' \int_{-\tau}^0 dt \, g^{(2)}(s,s',t',t) \right]\, .
\end{align}
It might be expected that, using this formula, the contribution from the slowest timescale would again be dominant for the MSE, because there are again integrals of an expression involving $g_N^{(1)}$ [recall \cref{g2_exact}]. 
However, it turns out that the situation is more complicated. First, there are a number of intermediate steps used to obtain Eq.~\eqref{eq:mse} that need to be reconsidered for the correlations given here. In particular, it is no longer valid to use the linear approximation from Ref.~\cite{Baker2020} that is used to derive Eq.~\eqref{eq:mse}. This is because there are large amplitude fluctuations associated with these multi-timescale statistics.

Details of the reconsidered derivation are given in Appendix \ref{app:bounds}, where we show that the linear approximation gives a qualitatively correct result, with the correction for amplitude fluctuations giving a logarithmic factor. Moreover, there is not just one optimal value of $\tau$ (the characteristic time for the phase estimate from heterodyne detection mentioned above). Surprisingly, the value of $\tau$ may be chosen to be any one of $N$ values, with each corresponding to one of the intrinsic timescales of the beam. 
Choosing $\tau$ to correspond to the $j^{\rm th}$ timescale (approximately $1/\ell_j$), the MSE in the linear approximation is then approximately
\begin{equation}\label{eq:mes_twoterms}
    \langle[\hat{\varphi}_F-\hat{\varphi}_R]^2\rangle\approx\frac 1{4\mathcal{N}_j \tau} + \ell_j \tau \, .
\end{equation}
In comparison, the ideal laser (a far-above-threshold amplifier) of Ref.~\cite{Baker2020}, which satisfies Glauber Ideality, differs from this expression only in that it has a factor of 2 on the second term (and has no $j$ subscript, of course). 
The resulting MSE for our model is then, to leading order,
\begin{equation}\label{eq:mse_ellenn}
    \langle[\hat{\varphi}_F-\hat{\varphi}_R]^2\rangle\approx\sqrt{\frac{\ell_j}{\mathcal{N}_j}} \approx \frac 2{\sqrt G}\, .
\end{equation}
That is, the minimum MSE behaves exactly as if we were measuring a beam from a {\em single} above-threshold amplifier. This situation is completely different from the coherence, which is strongly enhanced by the multiple timescales.

The case that is most comparable to a single amplifier is where $\tau$ is chosen to correspond to the fastest timescale (that of the final amplifier when the beam is produced by cascaded cavities). In that case, the parameters that are obtained in the expression for the MSE are the fastest decay rate $\ell_N$ and the overall output flux $\mathcal{N}_N$. For these beam statistics, this overall flux relates to $\ell_N$ in the same way, to leading order, as the flux of a single amplifier. That means the mathematical scaling for the MSE in Eq.~\eqref{eq:mse_ellenn} is the same as that for performing measurements directly on a single amplifier.

The next step in the derivation of the bound on the coherence is to use \cref{eq:MSE2} to compare the MSE for measurements on the output to the minimum circular variance for direct measurements on a quantum system with mean photon number $\mu$. In the relevant regime where phase error is small, we have $1-|\langle e^{i\left(\hat{\varphi}_F-\hat{\varphi}_R\right)}\rangle|^2\approx\langle[\hat{\varphi}_F-\hat{\varphi}_R]^2\rangle$ to leading order; this allows us to substitute \cref{eq:mse_ellenn} into \cref{eq:MSE2}, which gives
\begin{equation}
    \frac{4|z_A/3|^3}{\mu^2}\lesssim\frac 2{\sqrt G}.
\end{equation}
Even at this stage, the result is similar to that for a single amplifier.
What enables the larger bound on the coherence for $N>1$ timescales is that the coherence is much larger as a function of $G$ than for a single-timescale beam. Because the coherence for multiple timescales is proportional to $G^{N}$, the lower bound on $\mu$ is of the form
\begin{align}\label{lb_main_txt}
    \mu &\gtrsim \sqrt{2}\left|\frac{z_A}{3}\right|^{3/2}(\mathfrak{C}_N)^{1/4N} \, ,
\end{align}
rather than the lower bound $\propto\mathfrak{C}^{1/4}$ in \cref{eq:baker_ub} as follows from Ref.~\cite{Baker2020}.

Up to this point we have used the linear approximation to illustrate the similarities and differences from the single-amplifier case.
When accounting for the amplitude fluctuations, the MSE has a logarithmic factor, with the right-hand side of \cref{eq:mse_ellenn} replaced by approximately ${\ln G}/{\sqrt G}$.  (Again, see Appendix~\ref{app:bounds}.) That then gives the lower bound on $\mu$ as
\begin{align} \label{lbln}
    \mu &\gtrsim 2\left|\frac{z_A}{3}\right|^{3/2}\frac{\sqrt{N}(\mathfrak{C}_N)^{1/4N}}{(\ln\mathfrak{C}_N)^{1/2}} \, .
\end{align}
Thus, although the amplitude fluctuations do change the MSE, they only introduce a logarithmic factor in the bound we are able to derive. The main change in the bound, from that in \cref{eq:baker_ub}, is because the form of $g_N^{(1)}$ results in the integral giving $\mathfrak{C}$ scaling like $G^N$, not like $G$.

As expected, the model in the current paper, the cascaded amplifiers, respects the bounds \cref{lb_main_txt} and \cref{lbln}, with \cref{eq:asym_mu} having a right-hand side scaling as $(\mathfrak{C}_N)^{1/2N}$. It is also to be expected that our model does not attain the asymptotic lower bound scaling for $\mu$ as per \cref{lb_main_txt}. Our model employs standard linear couplings to produce a beam with the properties [Gaussian statistics with $g^{(1)}$ given by \cref{g1_approx}] used in deriving the lower bounds given here. Following the pattern of the result of Ref.~\cite{Baker2020} for ideal laser beams, one could speculate that an optimised quantum model, with arbitrary couplings, would be able to produce the same type of beam, with the same coherence, while having a quadratically smaller number of stored photons in the device. That would result in a saturation (in scaling) of \cref{lb_main_txt}. Whether that is possible is beyond the scope of this paper.

\section{Generalised analysis for two amplifiers}
\label{sec:realistic}

To this point, we have considered idealised parameter values, including identical intensity gains for each amplifier ($G_j=G$), arbitrarily large and identical timescale ratios between amplifiers ($r=\ell_j/\ell_{j-1}$), as well as arbitrarily many amplifiers in the cascade. In practice, there are of course limitations to how well these idealisations can be approximated. In this section, we therefore consider the $N=2$ case more carefully, using more general choices of parameters, as this is the simplest system that could give rise to a coherence improvement beyond a single amplifier. 

The parameter set we consider is
\begin{align}\label{eq:twocavityconfig}
\left\{G_1, \ell_1, G_2,r = \ell_2/\ell_1\right\},
\end{align}
and for these, we derive general expressions for the coherence and mean photon number of this system. Using these expressions, we first consider the behaviour of the coherence for asymptotically large $\mu$, and highlight the requirements on the timescale ratio $r$ for which a scaling improvement in coherence beyond that of a single amplifier can be expected. Following this, we analyse the behaviour of our system in the regime of moderate amplifier gains, which would be relevant to experimental platforms where there are limitations on the photon number within a given amplifier. Even with these limitations, we find that significant improvements in the coherence beyond that of a single amplifier (or an ideal standard laser model) are still possible.

From \cref{eq:coherencen}, we have
\begin{align}\label{eq:C_2}
    \mathfrak{C}_2 = G_1G_2 - 1,
\end{align}
and in Appendix~\ref{app:derive_g1}, we derive the total mean photon number in the two amplifier system, 
\begin{align}\label{eq:mu2}
    \mu = \frac{\left(\sqrt{G_1}-1\right)}{2}\left(1+\frac{\left(\sqrt{G_1}+1\right)\left(\sqrt{G_2}+1\right)}{(1+r)}\right) + \frac{\left(\sqrt{G_2}-1\right)}{2}.
\end{align}
In this expression, we see that $\mu$ is composed of the terms $(\sqrt{G_1}-1)/2$ and $(\sqrt{G_2}-1)/2$, which correspond to the mean photon number in the first and second amplifiers, respectively, if they were both driven by vacuum fields. There is also a term depending on $G_1$, $G_2$ and $r$, which corresponds to the additional photon number in the second amplifier resulting from it being driven by the first. The analysis that follows in this section will be centred around \cref{eq:C_2} and \cref{eq:mu2}. Note that these are independent of  the rate $\ell_1$, so there are effectively three relevant parameters here ($G_1$, $G_2$ and $r$).

\subsection{Large photon number regime}

To analyse the behaviour of the two-amplifier system in the large $\mu$ regime, we consider the constants $\xi>0$, $\zeta>0$, and $\sigma\geq0$. These constants are $\mathcal{O}(1)$ in $\sqrt{G_1}$, and defined such that
\begin{align}
    \xi = \sqrt{G_2/G_1}, \quad \quad r = \zeta G_1^{\sigma},
\end{align}
and hence $\mathfrak{C}_2 = \xi^2 G_1^2-1$. With these constants, large $\mu$ is obtained by taking $\sqrt{G_1}\gg1$, in which limit, \cref{eq:mu2} becomes
\begin{align}
    \mu = \frac{\left(1+\xi\right)}{2}\sqrt{G_1}+\frac{\xi G_1^{3/2}}{2(1+\zeta G_1^\sigma)}\left[1+\mathcal{O}\!\left(G_1^{-1/2}\right)\right]+ \mathcal{O}(1).
\end{align}
From these formulas for $\mu$ and $\mathfrak{C}_2$, we can identify four distinct cases for their asymptotic behaviour as $\sqrt{G_1}\to\infty$, depending on the chosen value of $\sigma$ (i.e., the scaling of $r$ relative to $G_1$):
\begin{equation}
\left\{
\begin{aligned}
&{\rm for}~\sigma = 0, 
&\quad&
\mu \sim \dfrac{\xi}{2(1+\zeta)} G_1^{3/2}
&\implies&
\mathfrak{C}_2 \sim \xi^2\left(\dfrac{2(1+\zeta)}{\xi}\mu\right)^{4/3},
\\
&{\rm for}~0 < \sigma < 1, 
&\quad&
\mu \sim \dfrac{\xi}{2\zeta} G_1^{3/2-\sigma}
&\implies&
\mathfrak{C}_2 \sim \xi^2\left(\dfrac{2\zeta}{\xi}\mu\right)^{\frac{4}{3-2\sigma}},
\\
&{\rm for}~\sigma = 1, 
&\quad&
\mu \sim \dfrac{1+\xi+\xi/\zeta}{2} G_1^{1/2}
&\implies&
\mathfrak{C}_2 \sim \dfrac{16\xi^2}{(1+\xi+\xi/\zeta)^4}\mu^{4},
\\
&{\rm for}~\sigma > 1, 
&\quad&
\mu \sim \dfrac{1+\xi}{2} G_1^{1/2}
&\implies&
\mathfrak{C}_2 \sim \dfrac{16\xi^2}{(1+\xi)^4}\mu^{4}.
\end{aligned}
\right.
\end{equation}

For a comparison, we consider these results against the coherence achieved by a single amplifier in the large $\mu$ regime
\begin{align}\label{eq:C_1}
    \mathfrak{C}_1 = 4\mu(\mu+1)\sim4\mu^2.
\end{align}
With $\sigma=0$, $\mathfrak{C}_2$ monotonically increases with both $\xi$ and $\zeta$; however, the scaling with $\mu$ is worse than that in $\mathfrak{C}_1$. For the case where $0<\sigma<1$, where $\mathfrak{C}_2\propto\mu^{4/(3-2\sigma)}$,  an improvement in the scaling beyond a single amplifier is possible for $\sigma>1/2$.
The prefactor monotonically increases with $\zeta$ for any choice of $\sigma$. With $\sigma<1/2$, this prefactor also monotonically increases with $\xi$, however this switches to a monotonic decrease for $\sigma>1/2$. This behaviour implies that with the choice $\sigma=1/2$, $\mathfrak{C}_2$ has identical scaling to $\mathfrak{C}_1$, but the prefactor for $\mathfrak{C}_2$ can be made arbitrarily larger by increasing $\zeta$ [while still ensuring $\zeta=\mathcal{O}(1)$ in $\sqrt{G_1}$]. For $\sigma=1$, the scaling of $\mathfrak{C}_2$ with $\mu$ is quadratically larger compared to $\mathfrak{C}_1$. The prefactor in this case is maximised for the choice $\xi = \zeta/(\zeta+1)$; this simplifies the formula for the coherence to $\mathfrak{C}_2\sim[\zeta/(\zeta+1)]^2\mu^4$, which increases monotonically with $\zeta$. In the case $\sigma>1$, the scaling of the coherence remains as $\mathfrak{C}_2\propto\mu^4$, however the prefactor for this scaling is larger for a given choice of $\xi$ compared to the $\sigma=1$ case. Choosing $\xi=1$ (corresponding to identical gains $G_1=G_2$) maximises the prefactor $16\xi^2/(1+\xi)^4$ to unity.

The case $\xi=1$ and $\sigma > 1$ corresponds exactly to the assumed regime of \cref{eq:asympCN} for $N=2$ (namely, $G_1=G_2$ and $r\gg G$), and the result is in agreement: 
\begin{align}
    \mathfrak{C}_2 \underset{r,\mu}{\sim} \mu^4.
    \,
\end{align}
This is the same power of $\mu$ as the Heisenberg limit for laser coherence from Ref.~\cite{Baker2020}. The prefactor here ($1$) is smaller than that in the bound 
derived there ($2.97$), but larger than the largest prefactor known for a laser model 
($0.224$)~\cite{Ostrowski2022a} even when one relaxes Condition \ref{cond:4.0} 
into one of ``Passably Ideal Glauber Coherence''~\cite{Ostrowski2022b}. 

\subsection{Moderate photon number regime}

Throughout this article, our analysis has revolved around how the coherence scales with the excitation number in the limit where the amplifier gain becomes arbitrarily large. While this is interesting from a fundamental perspective, it is somewhat removed from what is achievable in contemporary experiments. Phase-insensitive linear amplification near the quantum noise limit~\cite{Caves_2012} has been demonstrated on various platforms in both the optical~\cite{Tang_2004,Andersen_2007} and microwave~\cite{Bergeal_2010,Roch_2012,Xu_2023} domain. However, this performance is often limited to moderate photon numbers within the amplifiers due to various detrimental effects~\cite{Aumentado_2020}. 
Additionally, since optimal performance of our system requires the timescale ratio $r$ to greatly exceed the amplifier gains, this will impose further restrictions on available sizes of $G_j$. We therefore end this section with a brief analysis of our system in the regime where $G_j$ is moderate, and the excitation number in each amplifier is limited to only a few photons.

To benchmark the performance of our system, we will compare its coherence against two models. The first is that of a single amplifier, which is given by the exact (i.e., non-asymptotic) formula in \cref{eq:C_1} above. The second is the coherence of an idealised single-mode laser model with saturated linear gain, and linear coupling of the laser mode [annihilation operator $\hat{a}_{\rm las}(t)$ in the Heisenberg picture] to the continuum vacuum field [annihilation operator $\hat{v}(t)$] to produce the beam [annihilation operator $\hat{b}_{\rm las}(t) = \sqrt{\kappa_{\rm las}}\hat{a}_{\rm las}(t)+\hat{v}(t)$]~\cite{Louisell1973}. Neglecting technical noise and all other sources of damping, the dynamics of the laser mode can be described by a master equation $\dot\rho = \mathcal{L}\rho$ for its state matrix $\rho$. In the number basis, this equation takes the form~\cite{Wiseman_1999},
\begin{align}\label{eq:ideal_las}
    \frac{1}{\kappa_{\rm las}}\frac{d\rho_{m,n}}{dt} = \mu\left(\frac{2\sqrt{mn}}{m+n}\rho_{m-1,n-1}-\rho_{m,n}\right)-\frac{m+n}{2}\rho_{m,n} + \sqrt{(m+1)(n+1)}\rho_{m+1,n+1},
\end{align}
where $\rho_{m,n}:=\bra{m}\rho\ket{n}$ and $\kappa_{\rm las}$ is the output coupling rate of the laser mode to the beam. 

The coherence for the ideal laser model in \cref{eq:ideal_las}, $\mathfrak{C}_{{\rm las}}$, can be obtained by integrating the first-order Glauber coherence function for the beam. For the above model, described in the appropriate rotating frame, $g^(1)(t+\tau,t)$ is again real and $t$-independent in steady state, so the coherence is again given by  \cref{eq:coh_g1}. That is, in terms of the Liouvillian $\mathcal{L}$, we have
\begin{align}\label{eq:C_las}
        \mathfrak{C}_{\rm las} & = \int_{-\infty}^\infty d\tau\, \langle\hat{b}^\dagger_{\rm las}(t+\tau)\hat{b}_{\rm las}(t)\rangle_{\rm ss} \nn & = 2\kappa_{\rm las}\int_{0}^\infty d\tau\,{\rm Tr}\left[\hat{a}_{\rm las}^\dagger e^{\mathcal{L}\tau}\left(\hat{a}_{\rm las}\rho_{\rm ss}\right)\right] \nn & = -2\kappa_{\rm las}{\rm Tr}\left[\hat{a}_{\rm las}^\dagger \mathcal{L}_+^{-1}\left(\hat{a}_{\rm las}\rho_{\rm ss}\right)\right],
\end{align}
where $\mathcal{L}_+^{-1}$ is the inverse of $\mathcal{L}$ on its row space, and $\rho_{\rm ss}$ is the steady-state density operator. It is well known that in the limit of large excitation number $\mu$, the output beam produced by this model is well described by the ideal laser beam state described above in \cref{eq:id_g1}, with photon flux $\mathcal{N}_{\rm id} \sim \kappa_{\rm las}\mu$ and phase diffusion rate $\ell_{\rm id} \sim \kappa/2\mu$~\cite{Wiseman_1999}. In this limit, the correlation function is given by $\langle\hat{b}^\dagger_{\rm las}(t+\tau)\hat{b}_{\rm las}(t)\rangle \sim \kappa_{\rm las}\mu e^{-\kappa_{\rm las}|\tau|/4\mu}$, implying a coherence of $\mathfrak{C}_{\rm las} \sim 4\mathcal{N}_{\rm id}/\ell_{\rm id} = 8\mu^2$. This is a factor of two larger than the coherence of an ideal linear amplifier in the same limit. However, our analysis in this section is in the moderate photon number regime, for which there is no analytic expression for $\mathfrak{C}_{\rm las}$. We therefore compute this quantity numerically using the final line of \cref{eq:C_las}; in these calculations, the choice of $\kappa_{\rm las}$ is arbitrary and $\mathfrak{C}_{\rm las}$ depends only on the mean excitation number~\footnote{Strictly speaking, these numerical calculations also require a truncation of the Hilbert space of $\rho$. We truncate at a dimension of $30$, which ensures that the relative error in coherence is below $3\times10^{-9}$ for mean photon numbers $\mu<6$.}.

We define two performance ratios,
\begin{subequations}
    \begin{align}\label{eq:R1}
        \mathfrak{R}^{(1)}(G_1, G_2, r) := \frac{\mathfrak{C}_2(G_1, G_2)}{\mathfrak{C}_1\!\left(\mu(G_1,G_2,r)\right)},
    \end{align}
    \begin{align}\label{eq:Rlas}
        \mathfrak{R}^{\rm (las)}(G_1, G_2, r) := \frac{\mathfrak{C}_2( G_1, G_2)}{\mathfrak{C}_{\rm las}\!\left(\mu(G_1,G_2,r)\right)},
    \end{align}
\end{subequations}
which allow us to compare $\mathfrak{C}_2$ against the two models mentioned above for the same photon number. This is important to consider because $\mu$ is the essential resource that is used for the production of $\mathfrak{C}$ in the beam for a given model~\cite{Baker2020}. We achieve this by treating $\mathfrak{C}_1$ and $\mathfrak{C}_{\rm las}$ as implicit functions of $G_1$, $G_2$ and $r$, where $\mu$ for these models is chosen in accordance with \cref{eq:mu2} to ensure identical photon number between the models.

\begin{figure}[tbh]
\centering
\includegraphics[width=0.5\columnwidth]{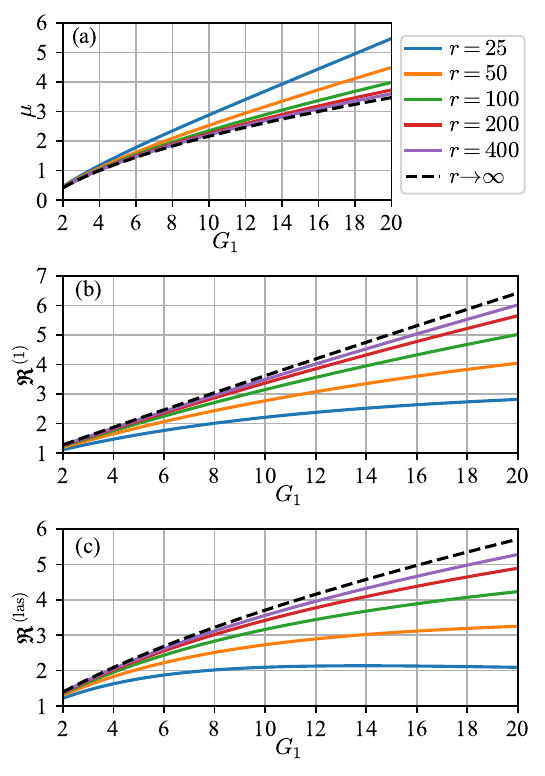}
\caption{\label{fig:2} (a) Photon number vs gain $G_1$ for a two-amplifier cascade [\cref{eq:mu2}].
(b) Performance ratio of the two-amplifier cascade and a single linear amplifier [\cref{eq:R1}].
(c) Performance ratio of the two-amplifier cascade and an ideal laser model [\cref{eq:R1}]. Here, we have set $G_2=G_1$, and different curves correspond to different choices for the value of $r$.
}
\end{figure}

In Fig.~\ref{fig:2}, we display the performance ratios $\mathfrak{R}^{(1)}$ and $\mathfrak{R}^{({\rm las})}$ by setting $G_1=G_2$ and modifying their behaviour by varying this gain parameter. Panel (a) plots the photon number $\mu(G_1,G_1,r)$ against $G_1$ for various choices of the bandwidth ratio $r$, while panels (b) and (c) show the performance ratios against $G_1$ for the same choice of values for $r$. Essentially, these plots reveal that $\mathfrak{C}_2$ can be substantially larger than both $\mathfrak{C}_1$ and $\mathfrak{C}_{\rm las}$ even for relatively modest amplifier gains, provided that $r$ remains moderately large. For example, with $G_1,G_2=10$ and $r=100$, we have $\mu\approx2.35$ ($\mu_1\approx1.08$, $\mu_2 \approx 1.27$), $\mathfrak{R}^{(1)}\approx3.15$, and $\mathfrak{R}^{({\rm las})}\approx3.16$. A further interesting point highlighted by these plots is that $\mathfrak{R}^{({\rm las})}$ can exceed $\mathfrak{R}^{(1)}$ for small $\mu$. This is because, while $\mathfrak{C}_{\rm las}\sim 2\mathfrak{C}_1$ in the limit of large $\mu$, $\mathfrak{C}_{\rm las}<\mathfrak{C}_1$  for $\mu$ less than about $2.3$.

To conclude this section, we present a final set of analytical results. We demonstrate it is possible to find an improvement in the coherence for a two-amplifier cascade over a single amplifier \textit{for any} choice of  $r>1$ (i.e., well outside the large $r$ regime) provided that $\mu$ is sufficiently large. Or, vice versa; an improvement can be found for any $\mu$ given a sufficiently large $r$. To show this, first note that taking $G_1 = 1$ in a two-amplifier cascade (no gain in the first amplifier) means that the second amplifier is fed with a vacuum field, so the resulting coherence is identical to just a single amplifier. The problem at hand then becomes a question of whether increasing $G_1$ above unity can yield a larger coherence for a given $\mu$ or $r$.

\begin{figure}[H]
\centering
\includegraphics[width=0.5\columnwidth]{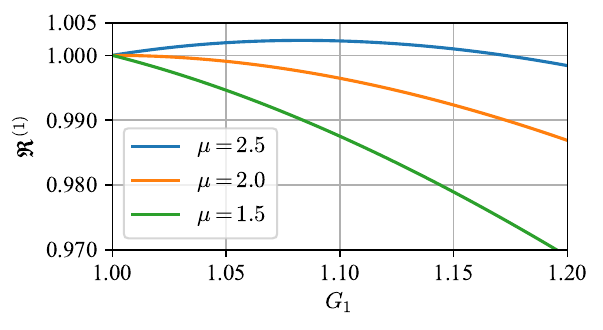}
\caption{\label{fig:3}Performance ratio [\cref{eq:R1_with_mu}] against gain $G_1$ with $r=2$. Three values for $\mu$ are chosen; $\mu=2$ is the threshold value given in \cref{eq:mu_thresh} for this value of $r$.
}
\end{figure}

To solve this problem, we first solve \cref{eq:mu2} for $G_2$ to give
\begin{align}\label{eq:G2_from_mu}
    G_2 = \left[\frac{\left(2\mu-\sqrt{G_1}+2\right)\left(1+r\right)-G_1+1}{G_1+r}\right]^2,
\end{align}
and then plug this expression into \cref{eq:R1}, so that
\begin{align}\label{eq:R1_with_mu}
    \mathfrak{R}^{(1)}(G_1,r,\mu) = \frac{G_1}{4\mu(\mu+1)}\left[\frac{\left(2\mu-\sqrt{G_1}+2\right)\left(1+r\right)-G_1+1}{G_1+r}\right]^2 - \frac{1}{4\mu(\mu+1)}.
\end{align}
In Fig.~\ref{fig:3}, we plot a few exemplary cases of \cref{eq:R1_with_mu} for $r = 2$. With the larger value of $\mu=2.5$, it is found that $\mathfrak{R}^{(1)}$ increases with $G_1$ up to a maximum, whereas for the smaller values of $\mu\in\{1.5,2.0\}$, the value of $\mathfrak{R}^{(1)}$ tends to monotonically decrease with $G_1$. This suggests a threshold value of $\mu$ for a given $r$, where above this threshold, having two amplifiers will give a coherence increase over a single one. 

To find this threshold, we solve the equation
\begin{align}\label{eq:deriv_R1}
    \left.\frac{d}{dG_1}\mathfrak{R}^{(1)}(G_1,r,\mu)\right|_{G_1=1}=0
\end{align}
for $\mu$. This gives the value at which the derivative of $\mathfrak{R}^{(1)}$ at $G_1=1$ changes from being negative to positive, where a positive (negative) derivative implies that $\mathfrak{R}^{(1)}>1$ ($\mathfrak{R}^{(1)}<1$) for some $G_1>1$. The only positive solution to \cref{eq:deriv_R1} is
\begin{align}\label{eq:mu_thresh}
    \mu_{\rm thresh} = \frac{2}{r-1}.
\end{align}
This means, for a given value of $r>1$ there can be an improvement in the coherence using two amplifiers for $\mu>\mu_{\rm thresh}$. Inverting Eq.~\eqref{eq:mu_thresh}, we also find $r_{\rm thresh} = 1+2/\mu$, which is the threshold value of $r$, given $\mu$, where an improvement in the coherence using two amplifiers is possible if $r>r_{\rm thresh}$.

Finally, it is worth highlighting from Eq.~\eqref{eq:mu_thresh} that $\mu_{\rm thresh}< 0$ whenever $r< 1$ (or infinite for $r=1$). This implies that it is impossible to obtain a coherence improvement ($\mathfrak{R}^{(1)}>1$) 
unless the second amplifier in the cascade is more broadband than the first. This fact can be shown in the following way.
First, from Eq.~\eqref{eq:mu2}, the condition $r< 1$ straightforwardly implies $\mu> (\sqrt{G_1G_2}-1)/2$. Under this condition, the coherence of a single amplifier becomes $\mathfrak{C}_1 = 4\mu(\mu+1)> G_1G_2-1 = \mathfrak{C}_2$, which then implies $\mathfrak{R}^{(1)}\leq 1$.

\section{Conclusion}
\label{sec:conc}

We have investigated limits to the production of coherence $\mathfrak{C}$ by a device with fixed mean photon number $\mu$. The coherence, or beam degeneracy,  is a fundamental property of bosonic fields that highlights a distinction between a laser beam and other common types of radiation (e.g., thermal). While an achievable upper bound scaling (i.e., Heisenberg limit) of $\mathfrak{C}\propto\mu^4$ exists for beams satisfying the Glauber Ideality Condition (Condition \ref{cond:4.0})~\cite{Baker2020}, we have shown that this can be greatly surpassed by dropping this condition on the properties of the beam. In particular, we demonstrated that a carefully chosen cascade of $N$ linear amplifiers can achieve a coherence $\mathfrak{C}_N\propto\mu^{2N}$.
A new lower bound on $\mu$ in terms of $\mathfrak{C}$ has also been derived for this system based on its beam statistics, which is at least quadratically smaller than the asymptotic relationship achieved by our system. This is as expected, since 
our device uses standard linear field couplings, and suggests that engineering optimal input-output relations of the device 
might allow a beam of the sort we consider to be produced with an even larger coherence for fixed $\mu$ and $N$. We also showed that, by allowing $N$ to increase without bound, a scaling of $\coh$ that is \emph{exponential} in $\mu$ is theoretically possible.

While much of this article has focused on the regime of asymptotically large photon number, we have also shown that significant coherence improvements can be achieved with the addition of a single amplifier (i.e., $N=2$), even at moderate photon numbers. 
Our initial results are encouraging and suggest no fundamental impediment to the system's realisation. However, a more systematic, model-specific feasibility analysis is needed to account for important detrimental processes such as cavity nonlinearities, pump depletion, additional loss, thermal noise, and other sources of technical imperfection.

Although achieving a high $\mathfrak{C}$, the cascaded system that we have investigated has beam properties that are very different from those of an ideal laser beam. This raises questions concerning the connection between coherence and the potential utility of our device as a phase reference. An ideal laser beam has one timescale (the coherence time, given by the reciprocal of its linewidth), while also exhibiting relatively low (Poissonian) intensity fluctuations. Conversely, our cascaded system has multiple very different coherence timescales, one per amplifier, so that $g_N^{(1)}(t)$ is approximately equal to a sum of $N$ decaying exponential functions, with their contribution to $g_N^{(1)}(0)=1$ getting successively larger as their coherence times get successively smaller.
Moreover, the unsaturated gain of each amplifier leads to large intensity fluctuations, with $g_N^{(2)}(0)=2$, the same as for thermal light. An interesting question for future work is whether modifying the system dynamics (considering above-threshold operation, for example) could provide more stable photon statistics, while maintaining high $\mathfrak{C}$.

The properties of an ideal laser beam are, of course, extremely useful within precision technology. However, there are particular applications where the high temporal and spatial coherence that is typical of laser radiation can give rise to coherent artefacts that can be detrimental to its performance. To avoid these effects, considerable effort has been directed towards the development of light sources with moderate temporal and/or spatial coherence~\cite{Lu_2024}. These sources have proven to be useful in areas including display and illumination~\cite{Redding_2012,Deng_2017}, optical coherence tomography~\cite{PODOLEANU_2012}, communication~\cite{Shen_2016} and random number generation~\cite{Li_2011}. With a high degeneracy $\mathfrak{C}$ but a multi-scale first-order Glauber coherence function $g_N^{(1)}(t)$, a system like the one we analyse here might be useful for these types of applications. 

As a final note, we return to a more fundamental question: if, as we have shown, the beam degeneracy $\coh$ is not, in general, bounded by a fixed power of $\mu$ (the mean photon number stored inside a device), is there some other property of the device for which this is true? 
Consider the regime of \cref{bothlarge}, where $\coh \sim (2\mu/N)^{2N}$, an unbounded power of $\mu$, is the scaling law resulting from our cascaded amplifier system. In this limit, each of the $N$ amplifier cavities is well approximated by an independent thermal state of photon number $\mu/N\gg1$. The von Neumann entropy for the total state of the device is thus, to leading order, $S_N = N[\ln(\mu/N)+1]$.
We can turn this entropy into an effective dimension of $D_{\rm eff}= e^{S_N}\approx (e\mu/N)^N$, from which we see $\mathfrak{C}\approx (2/e)^{2N}D_{\rm eff}^2$. This suggests there may be a universal upper bound for the coherence of a beam scaling as a power of the dimension of the device producing the beam. Here, by ``universal'' we mean regardless of the particular beam statistics, other than being unidirectional and stationary. 

Specifically, we conjecture that the ultimate quantum limit on $\mathfrak{C}$, for a beam and device satisfying only Conditions \ref{cond:1}--\ref{cond:3} of Ref.~\cite{Baker2020}, is given by $\coh = O(D^4)$, where $D$ is the effective Hilbert space dimension of the device. We note that Hilbert space dimension is a natural resource from the perspective of quantum information theory~\cite{Silva_2016,Plastino_2015,Brunner_2008,Sikora_2016}.
Our conjecture is consistent both with the result here and with the highest known coherence $\coh \sim 0.224\mu^4$ achievable from a single-mode non-standard cavity, in  Ref.~\cite{Ostrowski2022a}, where $D\sim 2\mu$.

\section*{Acknowledgements}
DWB worked on this project under a sponsored research agreement with Google Quantum AI.
This project is supported by Australian Research Council Discovery Projects DP210101367, DP220101602, and DP260102543.

\providecommand{\noopsort}[1]{}\providecommand{\singleletter}[1]{#1}%

\appendix

\section{Deriving the first-order Glauber coherence function and associated quantities}\label{app:derive_g1}

In this appendix, we evaluate several quantities of interest for our system in different parameter regimes. In Sec.~\ref{Apx:partA}, we restrict our attention to the case of identical gain (\(G_j=G\)) and timescale ratios (\(r = \ell_j/\ell_{j-1}\)), and the regime where $G\ll r$. This allows us to obtain simplified expressions for the normalised correlation function $g^{(1)}_{N}$, the photon flux $\mathcal{N}_N$, and the mean photon number $\mu$, for arbitrary $N$. These are relevant to the analysis presented in Sec.~\ref{sec:asymp}. In Sec.~\ref{Apx:partB}, we restrict our attention to the two-cavity case and evaluate the mean photon number $\mu$ for general $G_1$, $G_2$ and $r$. This is relevant to the analysis presented in Sec.~\ref{sec:realistic}.

We begin by obtaining a general expression for the correlation function $\langle\beta_{N}^{*}(t+\tau)\beta_{N}(t)\rangle$. This is determined by the inverse Fourier transform of the output power spectrum $\widetilde S_{N,\beta}(\omega)$ given in Eq.~\eqref{S_N,out_2}, which we reproduce here
\begin{equation}
\label{eq:S_N_beta_product}
2\pi \widetilde S_{N,\beta}(\omega) = \prod_{j=1}^{N} \left[ 1 + \Lorf_j(\omega) \right] - 1,
\end{equation}
where $\Lorf_j(\omega) = (\ell_j/2)^2(G_j-1)/[(\ell_j/2)^2 + \omega^2]$, using the parameters $\ell_j$ and $G_j$ defined in Eqs.~\eqref{eq:def_ellj} and \eqref{eq:def_Gj}. The product term, which we denote $P(\omega)$, can be expressed as a single rational function, and can therefore be expanded into a sum of simple fractions using partial fraction decomposition. Since the degree of the polynomial in $\omega^2$ is the same in the numerator and denominator of $P(\omega)$, the general form of the expansion is:
\begin{equation}
\label{eq:P_omega_partial_fraction}
P(\omega) = \prod_{j=1}^{N} \left[ 1 + \Lorf_j(\omega) \right] = 1 + \sum_{j=1}^{N} \frac{C_{j,N}}{(\ell_j/2)^2 + \omega^2},
\end{equation}
where the leading ``1'' is the result of polynomial long division. The coefficients $C_{j,N}$ are found systematically using the Heaviside method. To find a specific coefficient $C_{j,N}$, one multiplies the full expression $P(\omega)$ by the denominator of the $j$-th term, $[(\ell_j/2)^2 + \omega^2]$, and then evaluates the result at the value of $\omega^2$ that makes this denominator zero. This procedure isolates $C_{j,N}$ by causing all other terms in the sum to vanish. Specifically, we evaluate at $\omega^2 = -(\ell_j/2)^2$:
\begin{equation}
\label{eq:C_k_heaviside}
C_{j,N} = \left\{ \left[(\ell_j/2)^2 + \omega^2\right] P(\omega) \right\}_{\omega^2 = -(\ell_j/2)^2} = (\ell_j/2)^2(G_j-1) \prod_{k \neq j}^{N} \left(1+\frac{(G_k-1)}{1 - (\ell_j/\ell_k)^2}\right).
\end{equation}
Substituting this decomposition into the expression for the power spectrum simplifies it to a sum:
\begin{equation}
\label{eq:S_N_beta_sum}
\widetilde{S}_{N,\beta}(\omega) = \frac{1}{2\pi} \sum_{j=1}^{N} \frac{C_{j,N}}{(\ell_j/2)^2 + \omega^2}.
\end{equation}
The time-domain correlation function is then found by taking the inverse Fourier transform of $\widetilde{S}_{N,\beta}(\omega)$. By the linearity of the transform, we operate on each term individually. Using the standard Fourier transform pair for a Lorentzian function, we arrive at the final expression for the correlation function:
\begin{equation}
\label{eq:correlation_function_time_domain}
\langle\beta_N^*(t+\tau)\beta_N(t)\rangle = \mathcal{N}_Ng_{N}^{(1)}(t+\tau,t) = \sum_{j=1}^{N} \frac{C_{j,N}}{\ell_j} e^{-(\ell_j/2)|\tau|}.
\end{equation}

\subsection{Asymptotic results for identical gain and timescale ratios}
\label{Apx:partA}

In terms of the parametrisation given in Eqs.~\eqref{eq:ncavityconfig} and \eqref{eq:ncavityconfig2}, the factors comprising the product on the right-hand side of \cref{eq:C_k_heaviside} are
\begin{align}
    1 + \frac{(G_k-1)}{1 - (\ell_j/\ell_k)^2} = 1 + \frac{G-1}{1 - r^{2(j-k)}}.
\end{align}
In the regime where $G\ll r^2$, this term approaches $G$ for $k>j$ and approaches $1$ for $k<j$, so the product is dominated by $G^{N-j}$. This simplifies the form of the coefficient to
\begin{equation}
C_{j,N} = \frac{r^{2(j-1)}\ell^2_1(G-1) G^{N-j}}{4}\left\{1+\mathcal{O}\!\left(\frac{G-1}{r^2}\right)\right\}.
\end{equation}
Substituting this and the parametrisation $\ell_j = r^{j-1}\ell_1$ into the expression for the correlation function in Eq.~\eqref{eq:correlation_function_time_domain} yields
\begin{equation}
\label{eq:correlation_function_parameterised}
\mathcal{N}_Ng_{N}^{(1)}(t+\tau,t) = \frac{r^{N-1}\ell_1(G-1)}{4}\sum_{j=1}^{N}\left(\frac{G}{r}\right)^{N-j}e^{- r^{j-1}\ell_1|\tau|/2}\left\{1+\mathcal{O}\!\left(\frac{G-1}{r^2}\right)\right\}.
\end{equation}
Because this expression gives $\mathcal{N}_Ng_{N}^{(1)}(t+\tau,t)$, the value of $\mathcal{N}_N$ may be obtained by taking $\tau=0$. With $G\ll r^2$, we get
\begin{align}\label{N_more_general}
    \mathcal{N}_N = \frac{r^{N-1}\ell_1(G-1)}{4}\sum_{j=1}^{N}\left(\frac{G}{r}\right)^{N-j}\left[1+\mathcal{O}\!\left(\frac{G-1}{r^2}\right)\right].
\end{align}
This quantity reduces to a simpler expression if we further restrict the parameter regime, such that $G\ll r$. In this case, the $j=N$ term in the sum dominates, and we find
\begin{equation}
\label{eq:photon_flux}
\mathcal{N}_N=r^{N-1}\frac{\ell_1(G-1)}{4}\left[1+\mathcal{O}\!\left(\frac{G}{r}\right)\right] \sim \frac{\ell_N(G-1)}{4}.
\end{equation}
Dividing by this expression then gives the leading-order expression for $g_{N}^{(1)}(s,t)$ as
\begin{equation}
g_{N}^{(1)}(s,t)\approx\sum_{j=1}^{N}\left(\frac{G}{r}\right)^{N-j}e^{-r^{j-1}\ell_1|s-t|/2}.
\end{equation}

The restriction to the regime where $G\ll r$ is pertinent to our analysis in Sec.~\ref{sec:asymp}. This is because it minimises the value of $\mu$ with respect to $\mathfrak{C}_N$, for a given $G$ and $N$. To see this, note that for a given $G$ and $N$, $\mathfrak{C}_N$ in Eq.~\eqref{eq:CN} is independent of $r$, while $\mu$ in Eq.~\eqref{eq:photonnumber} is lower bounded by $\int d\omega \sum_{j=1}^NL_j(\omega)/2\pi\kappa_j = N(\sqrt{G}-1)/2$ due to the positivity of expressions within that equation. We will show that this lower bound is obtained asymptotically when $G\ll r$. 

From Eqs.~\eqref{S_N,out} and \eqref{eq:essjay}, we have
\begin{align}
    \widetilde{S}_{j,\alpha}(\omega) = \frac{L_j(\omega)}{2\pi\kappa_j} + \frac{1}{\gamma_j}\left(\widetilde{S}_{j,\beta}(\omega)-\widetilde{S}_{j-1,\beta}(\omega) - \frac{L_j(\omega)}{2\pi}\right) .
\end{align}
Using this in Eq.~\eqref{eq:mu_spec}, we get the exact expression
\begin{align}\label{mu_derivation}
\begin{split}
    \mu & = \int d\omega \sum_{j=1}^N\left[\frac{L_j(\omega)}{2\pi\kappa_j} + \frac{1}{\gamma_j}\left(\widetilde{S}_{j,\beta}(\omega)-\widetilde{S}_{j-1,\beta}(\omega) - \frac{L_j(\omega)}{2\pi}\right)\right] \\ & = \sum_{j=1}^N\left[ \frac{2\left(\mathcal{N}_j - \mathcal{N}_{j-1}\right)}{\ell_j\left(\sqrt{G_j}-1\right)} - 1\right].
\end{split}
\end{align}
Now setting $G_j=G$ and $\ell_j = r^{j-1}\ell_1$, we can use Eq.~\eqref{N_more_general} to evaluate the difference between the fluxes, where
\begin{align}
    \mathcal{N}_j - \mathcal{N}_{j-1} = \frac{r^{j-1}\ell_1(G-1)}{4}\left(1+\frac{G-1}{r}\sum_{k=1}^{j-1}\left(\frac{G}{r}\right)^{j-1-k}\right)\left[1+\mathcal{O}\!\left(\frac{G-1}{r^2}\right)\right].
\end{align}
Substituting this expression into the final line of \cref{mu_derivation}, we find
\begin{align}\label{mu_derivation_2}
\begin{split}
    \mu  = N\left[\frac{\sqrt{G}-1}{2} + \mathcal{O}\!\left(\left(\sqrt{G}+1\right)\frac{\left(G-1\right)}{r}\right) \right].
\end{split}
\end{align}
Taking $G\ll r$ in Eq.~\eqref{mu_derivation_2} gives $\mu= N(\sqrt{G}-1)/2$ to leading order; this is the lower bound for $\mu$ stated above. Therefore, $\mathfrak{C}_N$ is optimal with respect to $\mu$ in this limit for a given $G$ and $N$. This asymptotic expression for $\mu$ is reproduced in Eq.~\eqref{eq:mu_sim} of the main text.

\subsection{Mean photon number for two amplifiers}
\label{Apx:partB}

We now compute the mean photon number for $N=2$, with the general parameter set $\{G_1,\ell_1,G_2,r=\ell_2/\ell_1\}$. From \cref{mu_derivation}, we have
\begin{align}\label{eq:mu2_1}
    \mu = \frac{2\mathcal{N}_1}{\ell_1\left(\sqrt{G_1}-1\right)} + \frac{2\left(\mathcal{N}_2 - \mathcal{N}_1\right)}{r\ell_1\left(\sqrt{G_2}-1\right)} - 2.
\end{align}
Evaluating \cref{eq:correlation_function_time_domain} at $\tau=0$, we get
\begin{subequations}
    \begin{align}
        \mathcal{N}_1 = \frac{C_{1,1}}{\ell_1}= \frac{\ell_1(G_1-1)}{4} ,
    \end{align}
    \begin{align}
        \mathcal{N}_2 = \frac{C_{1,2}}{\ell_1}+\frac{C_{2,2}}{\ell_2}= \frac{\ell_1(G_1-1)}{4}\left(1+\frac{(G_2-1)}{1-1/r^2}\right)+\frac{r\ell_1(G_2-1)}{4}\left(1+\frac{(G_1-1)}{1-r^2}\right).
    \end{align}
\end{subequations}
Substituting these expressions into \cref{eq:mu2_1} and simplifying yields \cref{eq:mu2} in the main text.

\section{Bounds on coherence from phase estimation: heuristic derivation}
\label{app:bounds}

\subsection{Ideal Glauber coherence with negative exponential weighting}

Now we explain how the new Glauber coherence affects the derivation of the bound on the coherence given in the Supplementary Information of Ref.~\cite{Baker2020}.
That bound is derived by considering the accuracy of phase estimation that is possible based on measuring the output beam, and comparing it to the minimum error possible by measuring the cavity with mean photon number $\mu$.
There it is found that the mean-square error of phase estimation can be approximated by
\begin{align}
    \frac 1{\mathcal{N}\tau} &+ \frac 1{\tau^4} \left[ \frac 12 \int_0^\tau ds \int_{-\tau}^0 ds' \int_0^\tau dt' \int_{-\tau}^0 dt \, \delta g_{{\rm ideal}-1}^{(2)}(s,s',t',t) \right. \nn
    & \left. -\frac 12 \int_0^\tau ds \int_0^{\tau} ds' \int_{-\tau}^0 dt' \int_{-\tau}^0 dt \, \delta g_{{\rm ideal}-1}^{(2)}(s,s',t',t) \right]\, ,
\end{align}
where $\delta g_{{\rm ideal}-1}^{(2)}$ is the difference of $g_{{\rm ideal}}^{(2)}$ from $1$.
This expression suggests that the bound on the coherence should be primarily due to the second-order coherence, and so the larger coherence possible here should be due to the difference of $g^{(2)}$ from $g_{{\rm ideal}}^{(2)}$.

There are a number of steps used to derive Eq.~\eqref{eq:mse} in Ref.~\cite{Baker2020} that may no longer be valid, so here we provide an alternative derivation.
Following Ref.~\cite{Baker2020}, we define
\begin{align}
    \hat F &:= \int_{T-\tau}^T dt \, u_F(t)\hat b(t) + \hat a_F^\dagger \\
    \hat R &:= \int_{T}^{T+\tau} dt \, u_R(t)\hat b(t) + \hat a_R^\dagger \\
    \hat S &:= \hat R^\dagger \hat F \, ,
\end{align}
where $u_F(t)$ and $u_R(t)$ are normalised such that the integral of their squares are equal to 1.
In Ref.~\cite{Baker2020} the functions $u_F(t)$ and $u_R(t)$ were chosen such that they were constant over an interval of length $\tau$, and $\ell\tau\ll 1$.
That approach could be problematic when the system has correlations at multiple times, so we consider a more general case.

Taking the expectation value of $\hat S$ gives
\begin{align}
    \langle \hat S \rangle &= \int_{-\infty}^T dt \int_T^\infty ds \, u_F(t)u_R(s) \langle \hat b^\dagger(s) \hat b(t) \rangle \nn
    &= {\mathcal{N}} \int_{-\infty}^T dt \int_T^\infty ds \, u_F(t)u_R(s) g^{(1)}(s,t) \, ,
\end{align}
where we have extended the range of the integrals to infinity.
We can choose the functions $u_F(t),u_R(t)$ as
\begin{equation}
    u(t) = \frac 1{\sqrt{\tau}} \exp\left[ -\frac 1{2\tau} |t-T|\right] ,
\end{equation}
which is common for measurement of a varying phase.
Before evaluating the integrals for the correlations from multiple cavities, we consider the results for ideal Glauber coherence.

The ideal Glauber coherences are
\begin{align}
    g_{\rm ideal}^{(1)}(s,t) &= \exp\left[ -\frac{\ell}2 |s-t| \right]\, . \\
    g_{\rm ideal}^{(2)}(s,s',t',t) &= \exp\left[ -\frac{\ell}2 (|s-t|+|s'-t'|+|s-t'|+|t-s'|-|s-s'|-|t-t'|) \right] \, .
\end{align}
Then evaluating the integral yields
\begin{align}
    \langle \hat S \rangle &= \frac {\mathcal{N}}{\tau}  \int_{-\infty}^T dt \int_T^\infty ds \, \exp\left[ -\frac 1{2\tau} (|t-T|+|s-T|) -\frac{\ell}2 |s-t|\right] \nn
    &= \frac {\mathcal{N}}{\tau} \int_{-\infty}^T dt \int_T^\infty ds \, \exp\left[ -\left(\frac 1{2\tau} +\frac{\ell}2\right) (s-t)\right] \nn
    &= \frac {\mathcal{N}}{\tau} \left(\frac 1{2\tau} +\frac{\ell}2\right)^{-2} \nn
    &= \frac{4\mathcal{N}\tau}{(1+\ell\tau)^2}.
\end{align}
Note that this is approximately $4\mathcal{N}\tau$ in the case that $\ell\tau\ll 1$.
This means that the $\tau$ here is effectively equivalent to $1/4$ of the $\tau$ in Ref.~\cite{Baker2020}.
We could alternatively define $u(t)$ to eliminate the factor of $1/4$, but that would make later expressions more complicated.

Then Eq.~(S67) of Ref.~\cite{Baker2020} becomes (using $T=0$ because of time translation invariance)
\begin{align}
    \langle \hat S^\dagger \hat S \rangle &= 1 + \mathcal{N}\int_{-\infty}^0 ds \int_{-\infty}^0 dt \, u_F(s) u_F(t) g^{(1)}(s,t) + \mathcal{N}\int_0^{\infty} ds \int_0^{\infty} dt \, u_R(s) u_R(t) g^{(1)}(s,t) \nn
    & \quad + \mathcal{N}^2\int_0^\infty ds \int_{-\infty}^0 ds' \int_0^\infty  dt' \int_{-\infty}^0 dt \, u_R(s) u_F(s') u_R(t') u_F(t) g^{(2)}(s,s',t',t) \nn
&= 1+ \frac{8\mathcal{N}\tau}{1+\ell\tau} + \frac{16\mathcal{N}^2\tau^2}{(1+\ell\tau)^2}\nn
&= \frac{(1+\ell\tau+4\mathcal{N}\tau)^2}{(1+\ell\tau)^2}
\, .
\end{align}
Similarly, Eq.~(S72) of that work yields
\begin{align}
    \langle \hat S^2 \rangle &= \mathcal{N}^2\int_0^\infty ds \int_0^{\infty} ds' \int_{-\infty}^0  dt' \int_{-\infty}^0 dt \, u_R(s) u_R(s') u_F(t') u_F(t) g^{(2)}(s,s',t',t) \nn
    &= \frac{16\mathcal{N}^2\tau^2}{(1+\ell\tau)^2(1+2\ell\tau)^2}.
\end{align}
In the case where the fluctuations in $\hat S$ are small compared to its mean $\overline S$, Ref.~\cite{Baker2020} shows that the relevant mean-square error (MSE) can be approximated by
\begin{equation}\label{eq:minmse}
    \frac 12 \frac{\langle \hat S^\dagger \hat S \rangle-\langle \hat S^2 \rangle}{\overline S^2} .
\end{equation}
For the ideal Glauber coherences, that expression can be evaluated as
\begin{equation}
    \frac 12 \frac{\langle \hat S^\dagger \hat S \rangle-\langle \hat S^2 \rangle}{\overline S^2}
    = \frac{(1+\ell\tau)^3(1+2\ell\tau+8\mathcal{N}\tau)((1+\ell\tau)(1+2\ell\tau)+8\ell\mathcal{N}\tau^2)}{32\mathcal{N}^2\tau^2(1+2\ell\tau)^2} .
\end{equation}

We can also separate the expression for the MSE in a similar way as in Ref.~\cite{Baker2020}, with the first term corresponding to $g^{(1)}$ (from $\langle \hat S^\dagger \hat S \rangle$) being
\begin{equation}
    \frac 12 \left(1+ \frac{8\mathcal{N}\tau}{1+\ell\tau}\right) \left( \frac{(1+\ell\tau)^2}{4\mathcal{N}\tau}\right)^2
    = \frac{(1+\ell \tau)^3}{4\mathcal{N}\tau} + \mathcal{O}((\mathcal{N}\tau)^{-2}) \, ,
\end{equation}
and that corresponding to $g^{(2)}$ (from $\langle \hat S^\dagger \hat S \rangle$ and $\langle \hat S^2 \rangle$) being
\begin{equation}
    \frac 12 \left[\left( \frac{4\mathcal{N}\tau}{1+\ell\tau} \right)^2 - \frac{16\mathcal{N}^2\tau^2}{(1+\ell\tau)^2(1+2\ell\tau)^2}\right]
    \left( \frac{(1+\ell\tau)^2}{4\mathcal{N}\tau}\right)^2 = 2\ell\tau + \mathcal{O}((\ell\tau)^2) \, .
\end{equation}
Given that $\ell\tau$ is small, the expression for the first term is equivalent to that in Ref.~\cite{Baker2020} when we account for the fact that the $\tau$ here can be regarded as $1/4$ of that in Ref.~\cite{Baker2020}.
The second term is slightly different, though.
The optimal value of $\tau$ is
\begin{equation}
    \tau \approx \sqrt{\frac{1}{8\mathcal{N}\ell}} \, ,
\end{equation}
which gives a minimal value of the MSE
\begin{equation}
    \sqrt{\frac {2\ell}{\mathcal{N}}}+\mathcal{O}(\ell/\mathcal{N}) \, .
\end{equation}
The factor of $\sqrt 2 \approx 1.414$ is better than the factor of $\sqrt{8/3}\approx 1.633$ from Ref.~\cite{Baker2020}.
Up to this point, the only difference of this derivation from that in Ref.~\cite{Baker2020} is that we have used a negative exponential weighting $u_F(t),u_R(t)$ in the estimate, rather than using a fixed weight over some time interval.
With this negative exponential weight we obtain a slightly tighter bound on the coherence, but the result is otherwise unchanged.

\subsection{Bound for cascaded cavities: linear approximation}
\label{sec:minimising}

Next, we derive the MSE in the case of the cascaded cavities.
In this subsection we show the result using the linear approximation similar to Ref.~\cite{Baker2020}.
For the model we consider there are intensity fluctuations, which means the linear approximation is no longer valid.
In Appendix \ref{sec:correction} we show how to correct for the intensity fluctuations to derive the exact bound.

Because the model has Gaussian statistics,
\begin{align}\label{eq:g2Gauss}
    g^{(2)}(s,s',t',t) &= \mathcal{N}^{-2} \langle \hat b^\dagger (s) \hat b^\dagger(s') \hat b(t') \hat b(t)\rangle \nn
    &= \mathcal{N}^{-2} \langle \hat b^\dagger (s) \hat b(t)\rangle \langle \hat b^\dagger(s') \hat b(t')\rangle
    + \mathcal{N}^{-2} \langle \hat b^\dagger (s) \hat b(t') \rangle
    \langle \hat b^\dagger (s') \hat b(t) \rangle \nn
    &= g^{(1)}(s,t)g^{(1)}(s',t') + g^{(1)}(s,t')g^{(1)}(s',t) \, .
\end{align}
We can further use the results that $\hat F$, $\hat R$, $\hat F^\dagger$, and $\hat R^\dagger$ all mutually commute, and that $\langle \hat F \hat F\rangle=\langle \hat R \hat R\rangle=0$.
Assuming that $u_F$ and $u_R$ are symmetric also gives that $\langle \hat F^\dagger \hat F\rangle=\langle \hat R^\dagger \hat R\rangle$.
This then gives, using Wick's theorem
\begin{align}
    \langle \hat S^2 \rangle &= \langle \hat R^\dagger \hat F \hat R^\dagger \hat F \rangle \nn
    &= \langle \hat R^\dagger \hat F \rangle \langle \hat R^\dagger \hat F \rangle + \langle \hat R^\dagger \hat F \rangle \langle \hat R^\dagger \hat F \rangle \nn
    &= 2\overline S^2 \, ,
\end{align}
and
\begin{align}
    \langle \hat S^\dagger\hat S \rangle &= \langle \hat F^\dagger \hat R \hat R^\dagger \hat F \rangle \nn
    &= \langle \hat R^\dagger \hat F \rangle \langle \hat R^\dagger \hat F \rangle + \langle \hat R^\dagger \hat R \rangle \langle \hat F^\dagger \hat F \rangle \nn
    &= \overline S^2 + \langle \hat F^\dagger \hat F \rangle^2 \, .
\end{align}
We can also determine, using Eq.~\eqref{eq:g2Gauss},
\begin{equation}
    \langle \hat F^\dagger \hat F \rangle = 1+ \mathcal{N}\int_0^\infty ds \int_0^\infty  dt \, u_R(s)  u_R(t)  g^{(1)}(s,t) \, .
\end{equation}

For simplicity, in the following we will use $M:=\langle \hat F^\dagger \hat F \rangle$.
Then the linearised approximation in Eq.~\eqref{eq:minmse} would suggest that the variance should be
\begin{equation}
    \frac 12 \left( \frac{M^2}{\overline S^2} -1 \right).
\end{equation}
If we use
\begin{equation}
    g^{(1)}(s,t) \approx \sum_{j=1}^N\left(\frac{G}{r}\right)^{N-j} e^{-\ell_j |s-t|/2} \, ,
\end{equation}
for $\ell_j = r^{j-1}\ell_1$, then we get
\begin{align}\label{eq:sbar}
    \overline S &\approx \sum_{j=1}^N\left(\frac{G}{r}\right)^{N-j} \frac{4\mathcal{N}\tau}{(1+\ell_j\tau)^2} \, , \\
M &\approx 
1+\sum_{j=1}^N\left(\frac{G}{r}\right)^{N-j} \frac{4\mathcal{N}\tau}{1+\ell_j\tau}\, .
\label{eq:SS}
\end{align}
We aim to find the $\tau$ that gives the minimum of $M/\overline S$.

Now say that we take $\tau \propto 1/(\kappa r^{m-1})$, with $\kappa:=\kappa_1$.
For $\overline S$ the terms in the sum then scale as $r^{j}$ up to $j=m$, and beyond that they scale as $1/r^j$.
This implies that the dominant contribution to the sum is from $j=m$.
Similarly, for the sums in $M-1$ the terms will be increasing up until the point where $j=m$.
After that point the terms will decrease due to the factor of $G^{-j}$.
We first perform the analysis for $m=N$, in which case we include only terms in the sums from $j=N$, and the $+1$ for $M$, which gives
\begin{align}\label{eq:MSval}
    \frac{M}{\overline S}  \approx
    \frac{(1+\ell_N\tau)(1+\ell_N\tau + 4\mathcal{N} \tau)}{4\mathcal{N} \tau} \, .
\end{align}
This expression has a minimum for
\begin{equation}
    \tau = \frac {1}{\sqrt{\ell_N(\ell_N +4\mathcal{N})}}
    \propto \frac 1{\kappa r^{N-1}}
    \, .
\end{equation}
This value of $\tau$ satisfies $\tau \propto 1/(\kappa r^{m-1})$, consistent with the above assumption.
The resulting minimum value for $M/\overline S$ is
\begin{equation}\label{eq:paradox}
    \frac{\ell_N +2\mathcal{N}+\sqrt{\ell_N(\ell_N +4\mathcal{N})}}{2\mathcal{N}}
    =1+ \frac 12 \left( R+\sqrt{R(R+4)} \right)
    \, ,
\end{equation}
where 
\begin{equation}
    R := \frac {\ell_N}{\mathcal{N}} \approx 4/G \, .
\end{equation}
Thus the case $m=N$ gives
\begin{align}\label{eq:MSE_small_tau}
    \frac 12 \left( \frac{M^2}{\overline S^2} -1 \right) &\approx \frac{M}{\overline S} -1\nn
    &= \frac{2}{\sqrt{G}} + \frac 2G + \mathcal{O}(G^{-3/2}+r^{-1})
    \, .
\end{align}

Note that the leading-order terms in Eq.~\eqref{eq:MSval} give
\begin{equation}\label{eq:leading}
    \frac{M}{\overline S}-1 \approx \frac 1{4\mathcal{N} \tau} + \ell_N\tau \, .
\end{equation}
This expression has a simple interpretation as the MSE for measurement of a fluctuating phase with flux $\mathcal{N}$.
Using this expression would give the optimal $\tau$ of
\begin{equation}
    \tau \approx \sqrt{\frac 1{4\ell_N \mathcal{N}}},
\end{equation}
and
\begin{align}
    \frac 12 \left( \frac{M^2}{\overline S^2} -1 \right) \approx
     \frac{2}{\sqrt{G}} 
    \, .
\end{align}

In this analysis we have also included $+1$ for $M$, and the optimal value of $\tau$ is dependent on this term.
For $m<N$ the $j=m+1$ term is larger than $1$ so this term needs to be included rather than $+1$.
Including this term for $m<N$ gives
\begin{align}
    \frac{M}{\overline S}  \approx \frac{ \frac{4\mathcal{N}\tau}{1+\ell_m\tau} + \left(\frac{G}{r}\right)^{-1}\frac{4\mathcal{N}\tau}{1+\ell_{m+1}\tau}}
    {\frac{4\mathcal{N}\tau}{(1+\ell_m\tau)^2}} 
    &= 1+\ell_m\tau+\frac{r}{G}\frac{(1+\ell_m\tau)^2} {1+\ell_{m+1}\tau} \, .
\end{align}
This expression is minimised for
\begin{equation}
    \tau = \frac{1}{\ell_m \sqrt{G+1}} + \mathcal{O}(1/r) \, .
\end{equation}
which gives
\begin{equation}
    \frac{M}{\overline S} -1 = \frac{2}{\sqrt{G}}+\frac 2G + \mathcal{O}(G^{-3/2}+r^{-1})  \, .
\end{equation}
This expression is the same to leading order as the result found for $m=N$.
Note that the value of $\tau$ smaller by about a factor of $1/\sqrt G$ means that the $j=m$ and $j=m+1$ terms for $M$ are comparable.
Note also that the optimal $\tau$ being smaller than $1/\ell_m$ by a factor of $\sqrt G$ is consistent with the statement that $\tau$ corresponds to the $m^{\rm th}$ timescale of $1/\ell_m$, because the timescales are separated by factors of $r$ which is much larger than $G$.

The leading-order terms in this case are
\begin{equation}
    \frac{M}{\overline S} -1 \approx \frac 1{G\ell_m \tau} + \ell_m\tau \, .
\end{equation}
Note that the output flux from cavity $m$ is
\begin{equation}
    \mathcal{N}_m \approx r^{m-1}\frac{\ell_1 (G-1)}{4} \approx \frac{\ell_m G}{4}\, .
\end{equation}
Therefore the leading-order terms may be given as
\begin{equation}
    \frac{M}{\overline S} -1 \approx \frac 1{4\mathcal{N}_m \tau} + \ell_m\tau \, .
\end{equation}
This expression is equivalent to the MSE for measurement of the fluctuating phase output from cavity $m$.

\begin{figure}[tbh]
    \centering
    \includegraphics[width = 0.5\textwidth]{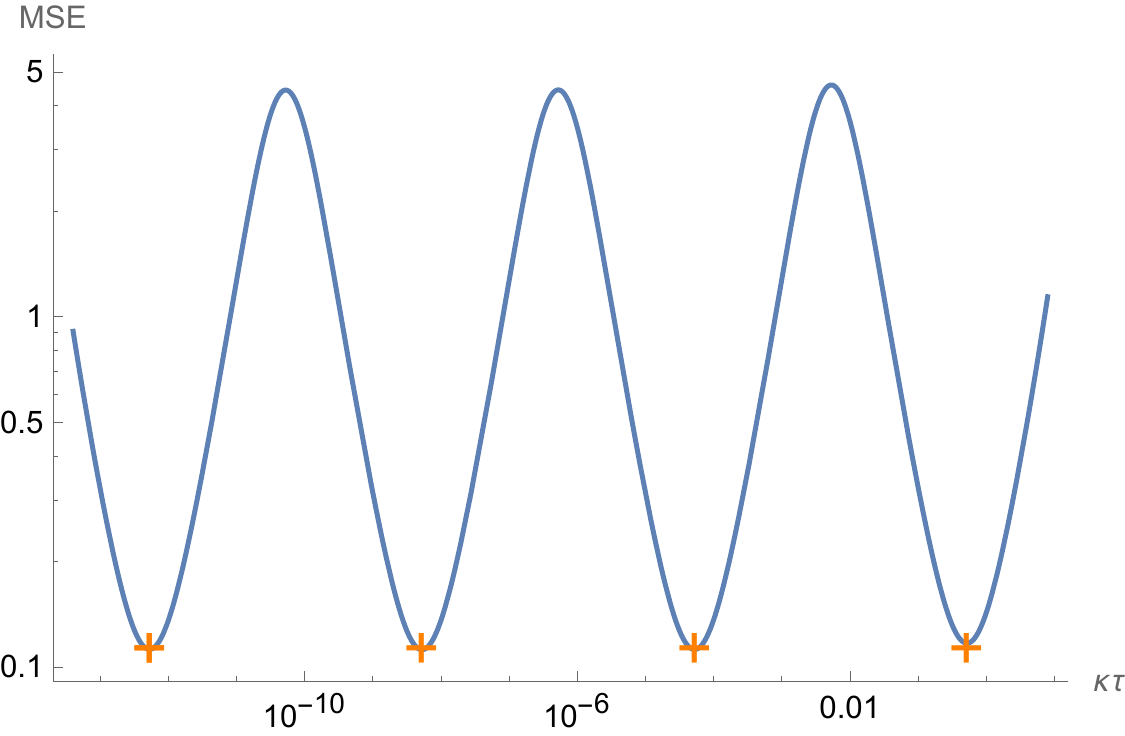}    
    \caption{The value of $((M/\overline S)^2-1)/2$ as a function of $\kappa\tau$, for $r=10^4$, $\gamma_j/\kappa_j=0.9$, and $N=4$.
The plusses show the predicted minima from theory, with the values of $m$ being 4 to 1 from left to right.
    }
    \label{fig:asymp}
\end{figure}

Thus this analysis predicts a series of $N$ local minima of similar size.
Numerical testing verifies this prediction.
As shown in Fig.~\ref{fig:asymp}, there are $N$ minima close to the predicted values.
The minimum for smallest $\tau$ is marginally smaller.
That may be expected because the others need two terms for $M$, though the analysis above finds the same leading-order terms. 

\subsection{Bound for cascaded cavities: exact}
\label{sec:correction}

The linearised approximation no longer holds here, but it is still possible to determine the approximation of the MSE by using the Gaussian properties of the field.
Because $\hat F$, $\hat R$, $\hat F^\dagger$, and $\hat R^\dagger$ all commute, they can be treated as Gaussian random variables in determining the expectation value
\begin{equation}\label{eq:expect}
    \left\langle \frac{\hat S}{\sqrt{\hat S^\dagger \hat S}} \right\rangle .
\end{equation}

Let us denote $\mathbf{v} := [F, R]^T$ for classical variables $F$ and $R$.
The probability density function for zero-mean correlated complex Gaussian variables is given by
\begin{equation}
    P(\mathbf{v}) = \frac{1}{\pi^2 \det(\Sigma)} \exp\left( -\mathbf{v}^\dagger \Sigma^{-1} \mathbf{v} \right) .
\end{equation}
The covariance matrix $\Sigma$ and its inverse are
\begin{equation}
    \Sigma = \begin{pmatrix} M & \overline S \\ \overline S & M \end{pmatrix}, \qquad 
    \Sigma^{-1} = \frac{1}{M^2 - \overline S^2} \begin{pmatrix} M & -\overline S \\ -\overline S & M \end{pmatrix}
\end{equation}
Let $D := \det(\Sigma) = M^2 - \overline S^2$, so the exponent becomes:
\begin{equation}
    -\mathbf{v}^\dagger \Sigma^{-1} \mathbf{v} = -\frac{M}{D}(|F|^2 + |R|^2) + \frac{\overline S}{D}(F^* R + R^* F) \, .
\end{equation}

We transform to polar coordinates, $F = r_1 e^{i\phi_1}$ and $R = r_2 e^{i\phi_2}$, so that
\begin{align}
    |F|^2 &= r_1^2, \qquad |R|^2 = r_2^2\, , \\
    F^* R + R^* F &= 2 r_1 r_2 \cos(\phi_2 - \phi_1)\, .
\end{align}
We define the phase difference $\theta := \phi_1 - \phi_2$, so the quantity we wish to calculate in Eq.~\eqref{eq:expect} is the expectation of the phase factor $e^{i\theta}$.
Due to the symmetry of the distribution ($\overline S$ is real), the imaginary part vanishes, and we can simply compute $\langle \cos \theta \rangle$.

The differential volume element for polar coordinates is $d^2F d^2R = r_1 dr_1 d\phi_1 \, r_2 dr_2 d\phi_2$.
We further change variables to the difference $\theta$ and sum $\Phi=\phi_1+\phi_2$.
The probability distribution is independent of $\Phi$, so we can integrate over $\Phi$ giving a factor of $2\pi$, and the expectation value to be calculated is
\begin{equation}
    \langle \cos \theta \rangle = \frac{2}{\pi D} \int_0^\infty \int_0^\infty \int_0^{2\pi} r_1 r_2 e^{-\frac{M}{D}(r_1^2+r_2^2)} e^{\frac{2\overline S}{D}r_1 r_2 \cos \theta} \cos \theta \, d\theta \, dr_1 \, dr_2 \, .
\end{equation}

We expand the interaction term involving $\cos \theta$ into a power series:
\begin{equation}
    e^{\frac{2\overline S}{D}r_1 r_2 \cos \theta} = \sum_{k=0}^{\infty} \frac{1}{k!} \left( \frac{2\overline S}{D} \right)^k (r_1 r_2)^k \cos^k \theta \, .
\end{equation}
Substituting this expression back into the integral, we separate the radial and angular parts
\begin{equation}
    \langle \cos \theta \rangle = \frac{2}{\pi D} \sum_{k=0}^{\infty} \frac{1}{k!} \left( \frac{2\overline S}{D} \right)^k \mathcal{I}^2_r(k) \mathcal{I}_\theta(k) \, ,
\end{equation}
where
\begin{align}
\mathcal{I}_r(k) &:= \int_0^\infty r^{k+1} e^{-\frac{M}{D} r^2} dr \, , \\
    \mathcal{I}_\theta(k) &:= \int_0^{2\pi} \cos^{k+1} \theta \, d\theta \, .
\end{align}

The angular integral is non-zero only when $k+1$ is even.
Let $k = 2j+1$, and the integral may be evaluated as
\begin{equation}
    \mathcal{I}_\theta(2j+1) = \frac{2\pi \binom{2j+2}{j+1}}{2^{2j+2}} \, .
\end{equation}
The integral in $r$ is evaluated as
\begin{equation}
    \int_0^\infty r^{k+1} e^{-\frac{M}{D}r^2} dr = \frac{1}{2} \left( \frac{D}{M} \right)^{\frac{k}{2}+1} \Gamma\left(\frac{k}{2}+1\right) .
\end{equation}
Putting these back into the expression for the expectation value yields
\begin{align}
    \langle \cos \theta \rangle &= \frac 2{\pi D} \sum_{j=0}^\infty \frac{1}{(2j+1)!} \left( \frac{2\overline S}{D} \right)^{2j+1} \frac{2\pi \binom{2j+2}{j+1}}{2^{2j+2}} \frac{1}{4} \left( \frac{D}{M} \right)^{2j+3} \Gamma^2\left(j + \frac{3}{2}\right)  \nn
    &= \frac{D}{2M^2} \sum_{j=0}^\infty \frac{1}{(2j+1)!} \left( \frac{\overline S}{M} \right)^{2j+1} \frac{(2j+2)!}{(j+1)!(j+1)!} \Gamma^2\left(j + \frac{3}{2}\right)  \nn
    &= \frac{D}{M^2} \sum_{j=0}^\infty  \left( \frac{\overline S}{M} \right)^{2j+1} \frac{1}{(j+1)!j!} \Gamma^2\left(j + \frac{3}{2}\right)  \nn
    &= \frac{\pi D}{4 M^2} \sum_{j=0}^\infty \frac{\Gamma^2(j+3/2)}{\Gamma^2(3/2)(j+1)!} \frac 1{j!} \left( \frac{\overline S}{M} \right)^{2j+1} \nn
    &= \frac{\pi D}{4M^2} \frac{\overline S}{M} \sum_{j=0}^\infty \frac{(3/2)_j (3/2)_j}{(2)_j} \frac 1{j!} \left( \frac{\overline S^2}{M^2} \right)^j \nn
    &= \frac{\pi D}{4M^2} \frac{\overline S}{M} \, {}_2F_1\left( \frac{3}{2}, \frac{3}{2}; 2; \frac{\overline S^2}{M^2} \right) \nn
    &= \frac{\pi}{4} \frac{\overline S}{M} \, {}_2F_1\left( \frac{1}{2}, \frac{1}{2}; 2; \frac{\overline S^2}{M^2} \right) \, ,
\end{align}
where ${}_2F_1$ is the hypergeometric function, and in the last line we have used Euler's hypergeometric transformation.

Now Eq.~(15.2.25) of Abramowitz and Stegun \cite{abramowitz_stegun} is
\begin{equation}
c(1-z){}_2F_1(a,b;c;z)-c{}_2F_1(a,b-1;c;z)+(c-a)z{}_2F_1(a,b;c+1;z)=0 \, .
\end{equation}
Taking $a=b=1/2$ and $c=1$ gives
\begin{equation}
(1-z){}_2F_1\left(\frac 12,\frac 12;1;z\right)-{}_2F_1\left(\frac 12,-\frac 12;1;z\right)+\frac 12 z{}_2F_1\left(\frac 12,\frac 12;2;z\right)=0 \, ,
\end{equation}
and so
\begin{align}
    {}_2F_1\left(\frac 12,\frac 12;2;z\right) &= \frac {2}{z}\left[ {}_2F_1\left(\frac 12,-\frac 12;1;z\right)- (1-z){}_2F_1\left(\frac 12,\frac 12;1;z\right) \right] \nn
    &= \frac {4}{\pi z}\left[ E(z) - (1-z) K(z) \right] \, .
\end{align}
Substituting $z = (\overline S/M)^2$ then gives
\begin{equation}
    \left\langle \frac{\hat S}{\sqrt{\hat S^\dagger \hat S}} \right\rangle = \frac{M}{\overline S} \left[ E\left(\frac{\overline S^2}{M^2}\right) - \left(1 - \frac{\overline S^2}{M^2}\right) K\left(\frac{\overline S^2}{M^2}\right) \right] \, .
\end{equation}

Then we can perform an expansion in $M^2/\overline S^2$ about 1 to obtain
\begin{equation}
    1-\left\langle \frac{\hat S}{\sqrt{\hat S^\dagger \hat S}} \right\rangle^2 = \frac 12 \left[ 4\ln 2 - 1 - \ln\left( \frac{M^2}{\overline S^2} - 1\right)\right]\left( \frac{M^2}{\overline S^2} - 1\right)
    + \mathcal{O}\left( \ln^2\left( \frac{M^2}{\overline S^2} - 1\right)\left( \frac{M^2}{\overline S^2} - 1\right)^2 \right).
\end{equation}
Thus we find that the expression without the linearised approximation yields a prefactor logarithmic in $M^2/\overline S^2-1$.
As shown in Appendix \ref{sec:minimising}, we can choose the filter such that
\begin{equation}
    \frac{M^2}{\overline S^2} - 1 \approx \frac 4{\sqrt G} + \mathcal{O}(G^{-1}) \, ,
\end{equation}
and so the MSE is
\begin{equation}
    2[2\ln 2 -1 + \ln \sqrt G]/\sqrt{G} + \mathcal{O}(G^{-1}\ln^2 G) \, .
\end{equation}
This minimum MSE would imply, following the reasoning in Ref.~\cite{Baker2020}, that
\begin{equation}\label{eq:lower}
    2[2\ln 2 -1 + \ln \sqrt G]/\sqrt G \gtrsim \frac{4|z_A/3|^3}{\mu^2} .
\end{equation}
Then we can use
\begin{align}
\mathfrak{C} &= G^N-1 \nn
    \mathfrak{C}^{1/N}&\approx G \, ,
\end{align}
so Eq.~\eqref{eq:lower} implies
\begin{align}
    \frac 2{\mathfrak{C}^{1/2N}} \left[2\ln 2 -1 + \frac 1{2N}\ln \mathfrak{C}\right]  &\gtrsim \frac{4|z_A/3|^3}{\mu^2} \nn
    \mathfrak{C} &\lesssim \frac{\mu^{4N}}{(4N)^{2N} |z_A/3|^{6N}} \ln^{2N}\mathfrak{C} \, .
\end{align}
Using $\mathfrak{C} \sim \mu^{4N}$ in the log above then gives
\begin{equation}
    \mathfrak{C} \lesssim \frac{(\mu^2\ln \mu)^{2N}}{|z_A/3|^{6N}}.
\end{equation}

\end{document}